\documentclass[camera,letterpaper,nomarginnotes,nonarrowgutter]{jpaper} 

\usepackage{fancyhdr}
\usepackage[normalem]{ulem}
\usepackage{balance}
\usepackage{bbding}
\usepackage{enumitem}
\usepackage{soul}

\usepackage{subcaption}
\usepackage{algorithmic}
\usepackage{graphicx,dblfloatfix}
\usepackage{textcomp}
\usepackage{xcolor}
\usepackage{xargs}                 

\usepackage{xspace}
\usepackage{xcolor}
\usepackage{pifont}
\usepackage{listings}
\usepackage{siunitx}
\usepackage{setspace}
\usepackage{multirow}
\usepackage{caption}
\usepackage{booktabs}

\newcommand{\revdel}[1]{}

\makeatletter
\g@addto@macro{\normalsize}{%
  \setlength{\abovedisplayskip}{3pt plus 0.5pt minus 1pt}
  \setlength{\belowdisplayskip}{3pt plus 0.5pt minus 1pt}
  \setlength{\abovedisplayshortskip}{0pt}
  \setlength{\belowdisplayshortskip}{0pt}
  \setlength{\intextsep}{4pt plus 1pt minus 1pt}
  \setlength{\textfloatsep}{4pt plus 1pt minus 1pt}
  \setlength{\skip\footins}{5pt plus 1pt minus 1pt}}
\makeatother

\usepackage{titlesec}
\titlespacing\section{0pt}{2pt plus 1pt minus 1pt}{3pt plus 1pt minus 2pt}
\titlespacing\subsection{0pt}{2pt plus 1pt minus 1pt}{3pt plus 1pt minus 2pt}
\titlespacing\subsubsection{0pt}{2pt plus 1pt minus 1pt}{3pt plus 1pt minus 2pt}

\newif\ifsubmission
\newif\ifrevision
\newif\iftacodraft
\newif\iftacorev
\submissiontrue
\revisionfalse
\tacodraftfalse
\tacorevtrue

\newcommand\X[0]{\textcolor{black}{Sectored DRAM\xspace}}

\newcommand\XMechOne[0]{\textcolor{black}{Sectored Activation\xspace}}
\newcommand\XMechTwo[0]{\textcolor{black}{Variable Burst Length\xspace}}
\newcommand\XMO[0]{\textcolor{black}{$SA$\xspace}}
\newcommand\XMT[0]{\textcolor{black}{$VBL$\xspace}}
\newcommand\XPredictor[0]{\textcolor{black}{Sector Predictor\xspace}}

\newcommand{\dingOne}{\ding{182}}
\newcommand{\dingTwo}{\ding{183}}
\newcommand{\dingThree}{\ding{184}}
\newcommand{\dingFour}{\ding{185}}

\newcommand{\squishlist}{
 \begin{list}{$\circ$}
  { \setlength{\itemsep}{0pt}
     \setlength{\parsep}{0pt}
     \setlength{\topsep}{0pt}
     \setlength{\partopsep}{0pt}
     \setlength{\leftmargin}{1em}
     \setlength{\labelwidth}{1em}
     \setlength{\labelsep}{0.5em} } }

\newcommand{\squishsublist}{
\begin{list}{$\rightarrow$}
 { \setlength{\itemsep}{0pt}
    \setlength{\parsep}{0pt}
    \setlength{\topsep}{-10em}
    \setlength{\partopsep}{-3pt}
    \setlength{\leftmargin}{1em}
    \setlength{\labelwidth}{1em}
    \setlength{\labelsep}{0.5em} } }

\newcommand{\squishend}{
  \end{list}  }

\ifsubmission
    \newcommand\outline[1]{}
    \newcommand\TODO[1]{}
        
    \newcommand{\new}[1]{#1}
    \newcommand{\atbc}[1]{}

    \newcommand{\nisc}[1]{}
    
    \definecolor{ao}{rgb}{0.0, 0.5, 0.0}
    
    \newcommand{\gfbox}[1]{}
    \newcommand{\gf}[1]{#1}
    
    \newcommand{\nis}[1]{#1}
    
    \newcommand\microrev[1]{#1}
    
    \newcommand{\tsub}[1]{#1}
    \newcommandx{\unsure}[2][1=]{}
    \newcommandx{\change}[2][1=]{}
    \newcommandx{\feedback}[2][1=]{}
    \newcommandx{\improvement}[2][1=]{}
    \newcommandx{\thiswillnotshow}[2][1=]{}
    \newcommandx{\completedRevision}[2][1=]{}
    \newcommandx{\dataSource}[2][1=]{}
    \newcommandx{\info}[2][1=]{}    
    
\else
\ifrevision
    \definecolor{gold}{rgb}{0.83, 0.69, 0.22}
    \definecolor{brightpink}{rgb}{1.0, 0.0, 0.5}
    \definecolor{ao}{rgb}{0.0, 0.5, 0.0}

    \newcommand\microrev[1]{\textcolor{red}{#1}}

    \newcommand{\tsub}[1]{\textcolor{black}{#1}}

    \newcommand{\new}[1]{\textcolor{black}{#1}}
    \newcommand{\atbc}[1]{}
        
    \newcommand{\nisc}[1]{}
    
    \newcommand{\gfbox}[1]{}
    \newcommand{\gf}[1]{\textcolor{black}{#1}}
    
    \newcommand{\nis}[1]{\textcolor{black}{#1}}
    
    \newcommand\loiscomment[1]{}
    \newcommand\agycomment[1]{}
    \newcommand{\hluocomment}[1]{}
    
    \newcommandx{\unsure}[2][1=]{\todo[linecolor=red,backgroundcolor=red!25,bordercolor=red,#1, size=\tiny]{#2}}
\newcommandx{\change}[2][1=]{\todo[linecolor=blue,backgroundcolor=blue!25,bordercolor=blue,#1,size=\scriptsize]{#2}}
\newcommandx{\feedback}[2][1=]{\todo[linecolor=yellow,backgroundcolor=yellow!25,bordercolor=yellow,#1]{#2}}
\newcommandx{\improvement}[2][1=]{\todo[linecolor=Plum,backgroundcolor=Plum!25,bordercolor=Plum,#1]{#2}}
\newcommandx{\thiswillnotshow}[2][1=]{\todo[disable,#1]{#2}}
\newcommandx{\completedRevision}[2][1=]{\todo[disable,backgroundcolor=red,#1]{#2}}
\newcommandx{\dataSource}[2][1=]{\todo[disable,backgroundcolor=red,#1]{#2}}
\newcommandx{\info}[2][1=]{\todo[linecolor=dollarbill,backgroundcolor=dollarbill!25,bordercolor=dollarbill,#1, size=\tiny]{#2}}
\else
\iftacodraft
    \newcommand\outline[1]{}
    \newcommand\TODO[1]{}
        
    \newcommand{\new}[1]{#1}
    \newcommand{\atbc}[1]{}

    \newcommand{\nisc}[1]{}
    
    \definecolor{ao}{rgb}{0.0, 0.5, 0.0}
    
    \newcommand{\gfbox}[1]{}
    \newcommand{\gf}[1]{#1}
    
    \newcommand{\nis}[1]{#1}
    
    \newcommand\microrev[1]{#1}
    \newcommand{\tsub}[1]{\textcolor{blue}{#1}}
    
    \newcommandx{\unsure}[2][1=]{}
    \newcommandx{\change}[2][1=]{}
    \newcommandx{\feedback}[2][1=]{}
    \newcommandx{\improvement}[2][1=]{}
    \newcommandx{\thiswillnotshow}[2][1=]{}
    \newcommandx{\completedRevision}[2][1=]{}
    \newcommandx{\dataSource}[2][1=]{}
    \newcommandx{\info}[2][1=]{}

\else 
    \definecolor{gold}{rgb}{0.83, 0.69, 0.22}
    \definecolor{brightpink}{rgb}{1.0, 0.0, 0.5}
    
    \newcommand\microrev[1]{\textcolor{red}{#1}}

    \newcommand{\tsub}[1]{\textcolor{blue}{#1}}
    \newcommand\outline[1]{\textcolor{cyan}{#1}}
    \newcommand\TODO[1]{\textcolor{magenta}{\textbf{TODO:} #1}}
        
    \newcommand{\new}[1]{\textcolor{blue}{#1}}
    \newcommand{\atbc}[1]{\textcolor{purple}{\textbf{ATB:} #1}}

    \definecolor{darkcyan}{rgb}{0.0, 0.55, 0.55}
    \newcommand{\nisc}[1]{\textcolor{darkcyan}{[\textbf{@nisa:} #1]}}
    
    \definecolor{ao}{rgb}{0.0, 0.5, 0.0}
    
    \newcommand{\gfbox}[1]{\textcolor{ao}{[\textbf{@GF:} #1]}}
    \newcommand{\gf}[1]{\textcolor{ao}{#1}}
    
    \newcommand{\nis}[1]{\textcolor{darkcyan}{#1}}
    
    \newcommandx{\unsure}[2][1=]{\todo[linecolor=red,backgroundcolor=red!25,bordercolor=red,#1, size=\tiny]{#2}}

\newcommandx{\change}[2][1=]{\todo[linecolor=blue,backgroundcolor=blue!25,bordercolor=blue,#1, size=\tiny]{#2}}
\newcommandx{\feedback}[2][1=]{\todo[linecolor=yellow,backgroundcolor=yellow!25,bordercolor=yellow,#1, size=\tiny]{#2}}
\newcommandx{\improvement}[2][1=]{\todo[linecolor=Plum,backgroundcolor=Plum!25,bordercolor=Plum,#1]{#2}}
\newcommandx{\thiswillnotshow}[2][1=]{\todo[disable,#1]{#2}}
\newcommandx{\completedRevision}[2][1=]{\todo[disable,backgroundcolor=red,#1]{#2}}
\newcommandx{\dataSource}[2][1=]{\todo[disable,backgroundcolor=red,#1]{#2}}
\newcommandx{\info}[2][1=]{\todo[linecolor=dollarbill,backgroundcolor=dollarbill!25,bordercolor=dollarbill,#1, size=\tiny]{#2}}

\fi
\fi
\fi

\newcommand{\tRP}{$t_{RP}$}
\newcommand{\tFAW}{$t_{FAW}$}

\newcommand\tempcommand[1]{\renewcommand{\arraystretch}{#1}}

\usepackage[colorinlistoftodos,prependcaption,textsize=small]{todonotes}
\usepackage{xargs}                    
\usepackage{titlesec}
\usepackage{titletoc}
\usepackage{clipboard}
\usepackage[all]{nowidow}
\usepackage[framemethod=tikz]{mdframed}
\usepackage[most]{tcolorbox}

\usepackage{marginnote}
\tcbuselibrary{breakable}
\tcbuselibrary{hooks}

\newcommandx{\changev}[2][1=]{}

\newcommandx{\finallabel}[2][1=]{}

\iftacorev
\newcounter{version}
\newcommand{\tacoreva}[1]{#1}
\definecolor{dark-green}{rgb}{0.00, 0.45, 0.00}
\newcommand{\tacorevb}[1]{#1}
\definecolor{goldbutdark}{rgb}{0.85, 0.65, 0.12}
\newcommand{\tacorevc}[1]{#1}
\newcommand{\tacorevd}[1]{#1}
\newcommand{\tacorevcommon}[1]{#1}

\newcommand{\tacofinalc}[1]{#1}

\definecolor{ao}{rgb}{0.0, 0.5, 0.0}
\definecolor{gerof}{rgb}{0.7, 0.0, 0.7}

\newcommand{\atbcr}[2]{#2}
\newcommand{\gercr}[2]{#2}

\newcommand{\omcr}[2]{#2}

\newcommand{\atbcrcomment}[1]{}
    \newcommand{\ignore}[1]{}
\fi

\newcommand\drambackgroundshort{\cite{ ghose2019demystifying, seshadri2019dram, kim2012case, zhang2014half,  seshadri2017ambit, lee2015adaptive, seshadri2013rowclone, Dennard68field,
keeth2008dram,
yauglikcci2022hira,
luo2023rowpress,oconnor2017fine,yuksel2024functionallycomplete,
bhati2015flexible,chang.sigmetrics2016,chang2017understanding,hassan2019crow,hassan2016chargecache,kim2018dram,kim2019d,kim2020revisiting,lee2017design,Tiered-Latency_LEE,lee2015decoupled,olgun2021quactrng}\xspace}

\usepackage{cleveref}

\crefformat{section}{\S#2#1#3}
\crefformat{subsection}{\S#2#1#3}
\crefformat{subsubsection}{\S#2#1#3}

\newcommand{\affilETH}[0]{\textsuperscript{\S}}
\newcommand{\affilETU}[0]{\textsuperscript{$\dagger$}}

\begin{document}
\bstctlcite{IEEEexample:BSTcontrol}
\title{{\X{}: A Practical Energy-Efficient and High-Performance Fine-Grained DRAM Architecture}}
\author{
{Ataberk Olgun\affilETH}\qquad%
{F. Nisa Bostanc{\i}\affilETH\affilETU}\qquad
{Geraldo F. Oliveira\affilETH}\qquad
{Yahya Can Tu\u{g}rul\affilETH\affilETU}\qquad
{Rahul Bera\affilETH}\qquad \\%
{A. Giray Ya\u{g}l{\i}kc{\i}\affilETH}\qquad%
{Hasan Hassan\affilETH}\qquad
{O\u{g}uz Ergin\affilETU}\qquad%
{Onur Mutlu\affilETH}\qquad\vspace{-3mm}\\\\
{\vspace{-3mm}\affilETH \emph{ETH Z{\"u}rich}} \qquad \affilETU \emph{TOBB University of Economics and Technology} %
}

\maketitle

\thispagestyle{plain}
\pagestyle{plain}
\begin{abstract}

\atbcr{1}{Modern computing systems access data in main memory at \emph{coarse} \emph{granularity} \atbcr{2}{(e.g., at 512-bit cache block granularity)}. Coarse\gercr{1}{-grained} access leads to wasted energy because the system does \emph{not} use all \gercr{1}{individually accessed small portions (e.g., \emph{words}, each of which typically is 64 bits) of a cache block}. In modern DRAM-based computing systems, two key coarse\gercr{1}{-grained} access mechanisms lead to wasted energy: large and fixed-size (i) data transfers between DRAM and the memory controller and (ii) DRAM row activations.}

We propose \X, a new, low-overhead DRAM substrate that \gf{reduces wasted energy} 
by enabling \emph{fine-grained} DRAM \atbcr{1}{data transfer} and \atbcr{1}{DRAM row} activation. 
To retrieve only useful data from DRAM, Sectored DRAM exploits the observation that many cache blocks are not fully utilized in many workloads due to poor spatial locality. Sectored DRAM predicts the words in a cache block that will likely be accessed during the cache block’s residency \atbcr{1}{in cache} and: (i) transfers only the predicted words on the memory channel by dynamically tailoring the DRAM data transfer size for the workload and (ii) activates a smaller set of cells that contain the predicted words by carefully operating physically isolated portions of DRAM rows (\omcr{1}{i.e., mats}). Activating a smaller set of cells on each access relaxes DRAM power delivery constraints and allows the memory controller to schedule DRAM accesses faster.

We evaluate \X{} using {41} workloads from widely-used benchmark suites. \new{Compared to a system with coarse-grained DRAM, }{\X{} reduce{s} the DRAM energy consumption of highly-memory-intensive workloads by up to (on average) {33\%} ({20\%}) while improving their performance by up to (on average) {36\%} ({17\%}). \X{}'s DRAM energy savings, combined with its system performance improvement, allows system-wide energy savings of up to {23\%}. \gercr{1}{\X{}'s DRAM chip area overhead is 1.7\% \atbcr{2}{of} the area of a modern DDR4 chip.} Compared to state-of-the-art fine-grained DRAM architectures, \X{} {greatly reduces DRAM energy consumption}, does \emph{{not}} reduce DRAM \atbcr{1}{bandwidth}, and can be implemented {with} low hardware cost. \atbcr{1}{\X{} provides 89\% of the performance benefits of, consumes 12\% less DRAM energy than, and takes up 34\% less DRAM chip area than a high-performance state-of-the-art fine-grained DRAM architecture (Half-DRAM).} {We hope and believe that \X{}'s ideas and results will help {to enable} more efficient and high-performance memory systems. {To this end, we open source} \X{} at \url{https://github.com/CMU-SAFARI/Sectored-DRAM}.}}

\end{abstract}

\sloppy
\section{Introduction}
\label{sec:introduction}

DRAM~\cite{Dennard68field} \new{is hierarchically organized}
\new{to improve} scaling
in density and performance. At the highest level \gf{of} the hierarchy, \revdel{\new{to enhance parallelism,}} \gercr{1}{a} DRAM \gercr{1}{chip is} partitioned into banks that can be accessed simultaneously~\cite{mutlu2008parallelism,kim2010thread,kim2012case,kim2016ramulator,lee2009improving}. 
At the lowest level, a \new{collection of {DRAM rows} \new{(DRAM cells that are activated together)} are} \gf{typically} divided into multiple \new{\emph{DRAM mats}} that can \new{operate} 
individually~\cite{keeth2008dram,itoh2013vlsi,vogelsang2010understanding,kim2012case}.\revdel{ This allows for easier routing of the wires that connect the DRAM cells to command and data buses and improves access latency by reducing capacitive loads on long wires~\cite{keeth2008dram,itoh2013vlsi,vogelsang2010understanding,kim2012case}. \new{DRAM mat organization provides} a useful basis for low-cost DRAM substrates that improve \gf{DRAM} energy efficiency.} 
Even though\revdel{ current } DRAM chips are hierarchically organized, standard DRAM interfaces (e.g., DDR\gf{x}~\cite{jedec2007ddr3,jedec2017ddr4,jedec2020ddr5}) do \emph{not} expose \new{DRAM mats}
to the memory controller. To access even a single DRAM cell, the memory controller \revdel{{first} }needs to activate a \gf{large number} of DRAM cells (\gf{e.g., 65,536 DRAM cells \new{in a DRAM row} in DDR4~\cite{micron2014ddr4}}
) and \revdel{then }transfer many bits (\gf{e.g.,} a cache block, typically 512 bits~\gf{\cite{hammarlund2014haswell}}) over the memory channel. {Thus, in current systems, \gf{both} DRAM \atbcr{1}{data transfer} and activation are \emph{coarse-grained}.} 
Coarse-grained \atbcr{1}{data transfer} and activation cause significant energy inefficiency in systems that use DRAM as main memory for two major reasons.

{First,\feedback{I think we can write this much cleaner. Define better the main source of the problem and its consequences.} coarse-grained \gf{DRAM} \atbcr{1}{data transfer}\revdel{ \new{(at cache block granularity)}} causes unnecessary data movement\revdel{ between the processor and DRAM}. \new{Standard DRAM interfaces transfer data at cache block granularity over fixed-size data transfer bursts (e.g., 8-cycle bursts in DDR4~\cite{jedec2017ddr4,micron2014ddr4}), but a large fraction of data \gercr{1}{(e.g., more than 75\%~\cite{qureshi2007line})} in a cache block is not used (i.e., referenced by CPU load/store instructions) during the cache block's residency in the cache hierarchy \new{(i.e., from the moment the cache block is brought to the on-chip caches until it gets evicted)}~\cite{kumar1998exploiting,kumar2012amoeba,pujara2006increasing,qureshi2007line,yoon2011adaptive,yoon2012dynamic}. Thus, transferring unused \gercr{1}{words} of a cache block over the power-hungry memory channel wastes energy~\cite{lee2017partial,zhang2017enabling,alawneh2021dynamic,son2014microbank,oconnor2017fine,chatterjee2017architecting,ha2016improving,udipi2010rethinking,paul2015harmonia,ware2010architecting,lefurgy2003energy}.}}

{Second, coarse-grained \gf{DRAM} activation causes an unnecessarily large \atbcr{1}{number} of DRAM cells \gercr{1}{in a DRAM row} to be activated. \new{Subsequent \gercr{1}{DRAM} accesses to \gercr{1}{\ignore{different cache blocks in} the activated row can}} be \new{served} faster.\ignore{ \gercr{1}{because accessing an activated DRAM row is faster than accessing a closed (i.e., not activated) DRAM row}.} \gercr{1}{However,} \new{many} modern memory-intensive workloads with irregular access patterns cannot benefit from \gercr{1}{these} fast\gercr{1}{er row} accesses as the spatial locality in these workloads is lower than the DRAM row size~\cite{lee2017partial,ghose2019demystifying,mutlu2013memory,mutlu2007memory,mutlu2007stall,sudan2010micro,yuan2009complexity,nesbit2006fair,subramanian2016bliss}. Thus, the energy cost of activating all cells in a DRAM row is not amortized over many accesses to the same row, leading to energy waste from activating a disproportionately large \atbcr{1}{number} of cells.}

Prior works~\cite{lee2017partial,zhang2014half,zhang2017enabling,alawneh2021dynamic,son2014microbank,oconnor2017fine,chatterjee2017architecting,cooper2010fine,ha2016improving,udipi2010rethinking} develop DRAM substrates that enable \emph{fine-grained} \gf{DRAM \atbcr{1}{data transfer} and} activation\new{, allowing \gercr{1}{words} of a cache block to be individually retrieved from DRAM and a small number of DRAM cells to be activated with each DRAM access}\revdel{ eliminate the energy wasted by activating unnecessarily large amount of DRAM cells}. However, \gf{these prior works 
(i) cannot provide high DRAM throughput~\cite{cooper2010fine,udipi2010rethinking}, 
(ii) incur high DRAM area overheads~\cite{zhang2014half,ha2016improving,zhang2017enabling,alawneh2021dynamic,son2014microbank,oconnor2017fine,chatterjee2017architecting}, and 
(iii) do \emph{not} fully enable\gercr{1}{\footnote{\gercr{1}{This class of prior works either do \emph{not} enable fine-grained data transfer (i.e., they \atbcr{2}{perform} data transfers at cache block granularity)~\cite{zhang2014half,cooper2010fine,udipi2010rethinking,ha2016improving} or do \emph{not} enable fine-grained data transfer for both read and write operations~\cite{lee2017partial}.}}} fine-grained DRAM~\cite{cooper2010fine,udipi2010rethinking,zhang2014half,ha2016improving,lee2017partial} (\cref{sec:motivation:limitations}).}

\revdel{fine-grained activation alone cannot eliminate the energy inefficiency caused by coarse-grained DRAM access. {\new{Moreover, these works} have at least one of the \new{three} shortcomings}{: they (i) propose intrusive modifications to DRAM array circuit and organization~\cite{oconnor2017fine,zhang2017enabling,son2014microbank,alawneh2021dynamic,chatterjee2017architecting}, (ii) greatly reduce the throughput of DRAM data transfers~\cite{udipi2010rethinking,cooper2010fine}, (iii) introduce considerable DRAM chip area overhead~\cite{zhang2014half,ha2016improving}.} \new{One prior DRAM substrate~\cite{lee2017partial} enables \emph{fine-grained DRAM activation} and a limited form of \emph{fine-grained DRAM access}, allowing the processor to \new{write to DRAM at} word granularity \new{(e.g., 8 bytes)} \new{instead of cache block granularity}.
%
Because this substrate does \emph{not} allow \gf{the processor} \new{to read from DRAM} at word granularity,
it does \emph{not} fully exploit fine-grained DRAM access (\cref{sec:motivation:limitations}).\revdel{ We show that fully exploiting fine-grained DRAM access is critical to eliminate energy waste and thus justify the costs of enabling fine-grained DRAM access (\cref{sec:results-perf-energy}).}}\footnote{No prior work develops a DRAM substrate that enables only fine-grained DRAM access.}
}

Our \textbf{goal} is to develop a new, low-cost, and high-throughput DRAM substrate that can mitigate the excessive energy consumption from both (i) {transmitting unused data on the memory \new{channel}} and (ii) {activating {a disproportionately large \atbcr{1}{number}} of DRAM cells}. \unsure{No key idea?}{To this end, we develop \omcr{1}{\emph{\X{}}}.} \atbcr{1}{\X{} leverages two key ideas to enable fine-grained data transfer and row activation at low chip area cost.}
\atbcr{1}{First, a cache block transfer between main memory and the memory controller happens in \emph{a fixed number of} \gercr{1}{DRAM interface} clock cycles where only a word of the cache line is transferred in each cycle. \X{} augments the memory controller and the DRAM chip to \atbcr{2}{perform} cache block transfers in a \emph{variable number of} clock cycles based on \gercr{1}{the} workload access pattern.} \atbcr{1}{Second, a large DRAM row, by design, is \emph{already} partitioned into smaller independent physically isolated regions. \X{} provides the memory controller with the ability to activate each such region based on \gercr{1}{the} workload access pattern.}

\X{} implements (i) \XMechTwo{} (\XMT{}) to enable fine-grained \gf{DRAM} \atbcr{1}{data transfer}, and \XMechOne{} (\XMO{}) to enable fine-grained \gf{DRAM} activation.
\revdel{\X{} eliminates the need to transfer unused data over the memory channel by effectively enabling control over DRAM access granularity using \XMT{}. \XMT{} allows the memory controller to send and receive \emph{dynamically-sized} portions of a cache block over the memory \new{channel}. \XMT{} leverages the key observation that DRAM data transfers occur over multiple \gercr{1}{DRAM interface} cycles in a burst. In each cycle, a \gercr{1}{word} \gercr{1}{(typically 64 bits)} of the cache block is transferred.}
{\XMT{} dynamically adjusts the number of cycles in a burst to transfer a different word of a cache block with each DRAM interface {cycle}, thus enabling fine-granularity DRAM \atbcr{1}{data transfer}.} \new{To do so at low cost, \XMT{} builds on existing DRAM I/O circuitry that already\revdel{implements a mechanism to} select\gf{s} one \gercr{1}{word} of a cache block to transfer in one cycle of a burst.}\revdel{ \new{Deciding which portions of a cache block to retrieve from DRAM is critical for maintaining high system performance. Portions that are not fetched but later referenced by load/store instructions cause additional high-latency memory accesses, potentially degrading system performance.}}

To enable \XMO{} with low hardware cost,\revdel{ {\X{} builds on the structures that already exist in the lowest level of the DRAM organization hierarchy.} {We make the key observation}} \gf{we leverage the fact} that DRAM {rows are already partitioned into \tsub{independent physically isolated \gercr{1}{regions}} \new{(mats)}} that can be individually activated with small modifications to the DRAM \new{chip}. \new{We refer to \tsub{a mat} that \tsub{incorporates these modifications} as a \emph{sector}.}\footnote{\new{We use the word \emph{sector} to distinguish between what exists today in DRAM chips (mats) and what we propose in \X{} (sectors).}} 
\tsub{Activating a sector consumes considerably smaller energy than activating a DRAM row as a sector typically contains \omcr{1}{almost} an order of magnitude fewer cells (e.g., 1024 in a DDR4 chip) than a DRAM row (e.g., typically 8192 in a DDR4 chip).} 
{}\XMO{} (i) implements \emph{sector transistors} that are each turned on to activate one of the independent \new{mats}, and (ii) \emph{sector latches} that control the sector transistors. \XMO{} exposes the sector latches to the memory controller by \gercr{1}{using} \new{an} existing DRAM command \gf{(\cref{sec:sectored-activation-mechanism})}, therefore, \gercr{1}{\XMO{} can be implemented without any changes to the physical DRAM interface.} As the power required to activate a \new{mat} in a DRAM row is only a fraction of the power required to activate the whole row, \X{} also relaxes the power delivery constraints in DRAM chips~\cite{lee2017partial,zhang2014half,oconnor2017fine}. \omcr{1}{Doing so} allows for the activation of DRAM rows at a higher rate, increasing memory-level parallelism for {memory-intensive workloads\revdel{ that have irregular access patterns}}. 

\atbcr{1}{\XMT{} and \XMO{} provide two key primitives for power-efficient, sub-cache-block-sized (e.g., 8-byte or \atbcr{1}{one-word}) data transfers between main memory and the rest of the system. However, because modern systems are typically designed to have cache-block-sized (e.g., 64-byte) data transfers between system components, making performance- and energy–efficient use of two \X{} primitives (\XMT{} and \XMO{}) requires system-wide modifications in hardware.} 
{We develop two \atbcr{1}{hardware} techniques (\cref{sec:accurate-word-retrieval}), (i) Load/Store Queue (LSQ) Lookahead and (ii) \XPredictor{} (SP) to effectively integrate \gf{\X{}} into a system. \gf{\atbcr{1}{At a high level, LSQ Lookahead and SP determine and predict, respectively,} which \gercr{1}{words} of a cache block \atbcr{1}{should be} retrieved from \atbcr{1}{a lower-level component of the memory hierarchy}.} 
\atbcr{1}{Accurately determining the words of a cache block that \gercr{1}{are} used during the cache block's \emph{residency} in system caches enables} high system performance and \atbcr{1}{low system energy consumption by improving data reuse in system caches as opposed to repeating a high-latency main memory access for each used word of a cache block}.

LSQ Lookahead accumulates the individual words in a cache block accessed by younger load/store instructions in older load/store instructions' memory requests. Thus, the execution of a load/store instruction prefetches the portions of cache blocks that will be accessed by the in-flight (i.e., not yet executed) load/store instructions. SP predicts which portions of a cache block will be accessed by a load/store instruction based on that instruction's past cache block usage patterns. This allows SP to accurately predict the portions of a cache block that will be used by the processor during the cache block's residency in the cache hierarchy.

\Copy{R3min/2a}{We evaluate the performance and energy of \X{} using 41 workloads from SPEC2006~\cite{spec2006} and 2017~\cite{spec2017} and DAMOV~\cite{oliveira2021damov, damov.github} benchmarks \atbcr{1}{using Ramulator~\cite{ramulator.github,ramulator2github,kim2016ramulator,luo2023ramulator}, DRAMPower~\cite{chandrasekar2012drampower}, and Rambus Power Model~\cite{rambuspowermodel}}. \finallabel{\ref{q:r3q2}}\X{} significantly reduces system energy consumption and improves system performance for \tacofinalc{memory-intensive workloads with irregular access patterns (which amounts to 10 of our workloads)}. \tacofinalc{For such workloads,} \new{compared to a system with \omcr{2}{conventional} coarse-grained DRAM,} \X{} reduces DRAM energy consumption by {20\%}, improves system performance by {17\%}, and reduces system energy consumption by {14\%}, \new{on average}. \atbcr{1}{\X{} does so as it} 1) improves workload execution time by issuing activate commands at a higher rate and thereby reducing average memory latency and 2) activates fewer DRAM cells and retrieves fewer sectors from DRAM at lower power. We estimate the DRAM area overheads of \X{} using CACTI~\cite{cacti} and find that it can be implemented with low hardware \gercr{1}{cost}. \X{} incurs {0.39 $mm^2$} DRAM area overhead ({1.7\%} of a \gf{DRAM} chip) and \new{does \emph{not} require modifications to the physical DRAM interface.} \atbcr{1}{Compared to the evaluated state-of-the-art fine-grained DRAM architectures~\cite{cooper2010fine,zhang2014half,lee2017partial,yoon2012dynamic}, \X{} {greatly reduces DRAM energy consumption}, does \emph{{not}} reduce DRAM \atbcr{1}{bandwidth}, and can be implemented {with} low hardware cost. \atbcr{1}{\X{} provides 89\% of the performance benefits of, consumes 12\% DRAM energy than, and takes up 34\% less DRAM chip area than a high-performance state-of-the-art fine-grained DRAM architecture~\cite{zhang2014half}. \X{} provides 10\% larger performance  and 13\% larger DRAM energy benefits compared to a low-cost state-of-the-art fine-grained DRAM architecture~\cite{lee2017partial}.}} \tacorevcommon{We open source our simulation infrastructure and all datasets to enable reproducibility and help future research~\cite{self.github}.}}

We make the following contributions:

\begin{itemize}[noitemsep,topsep=0pt,parsep=0pt,partopsep=0pt,labelindent=0pt,itemindent=0pt,leftmargin=*]

\item We \new{introduce} \X{} \new{and its two key} mechanisms\omcr{1}{:} \XMechTwo{} and \XMechOne{}. \X{} \new{improves system performance} and alleviates \gercr{1}{system energy consumption} by enabling \new{fine-grained DRAM \omcr{1}{data transfer} and activation.}

\item We \new{develop two techniques \new{(Load Store Queue Lookahead and Sector Predictor)} to effectively integrate \X{} into a system}. 
    \new{Our techniques} reduce the number of \gercr{1}{high-latency memory accesses by accurately} identifying the \gercr{1}{words} of a cache block that will be used by the processor.

\item We evaluate \X{} \atbcr{1}{with a wide range of workloads and observe that it provides higher system performance and energy efficiency than coarse-grained DRAM} \omcr{2}{as well as multiple prior fine-grained DRAM proposals}.

\item \atbcr{1}{We open source \X{} at \url{https://github.com/CMU-SAFARI/Sectored-DRAM}.}
\end{itemize}

\section{DRAM Background}
\label{sec:background}

We provide \atbcr{1}{the most} relevant background on DRAM organization \atbcr{1}{for our work}.\footnote{We refer the reader to {various} prior works~{\drambackgroundshort}~for a more detailed description of the DRAM architecture.}

\subsection{DRAM Organization}
\label{sec:background:dramorg}

\atbcr{1}{A typical computing system implements a \emph{memory controller} in the processor chip. The memory controller connects to multiple \emph{DRAM modules} over multiple \emph{memory channels}.}
\gf{Fig.}~\ref{fig:dram_organization} illustrates the hierarchical DRAM organization \atbcr{1}{inside a DRAM module (a)}. 
Multiple \emph{DRAM chips} \atbcr{1}{(b)} constitute a DRAM module. All \gf{DRAM} chips \atbcr{1}{\omcr{2}{in} a module} operate in lockstep \atbcr{1}{where they receive the same DRAM commands from the memory controller at the same time and respond to commands in unison}~\cite{kim2019d,chang2017understanding}. A \gf{DRAM} chip \gf{has} multiple \emph{DRAM banks} (c) that can be accessed in parallel. All banks in a chip share \atbcr{1}{an input/output logic}.

\begin{figure}[!ht]
    \centering
    \includegraphics[width=1\linewidth]{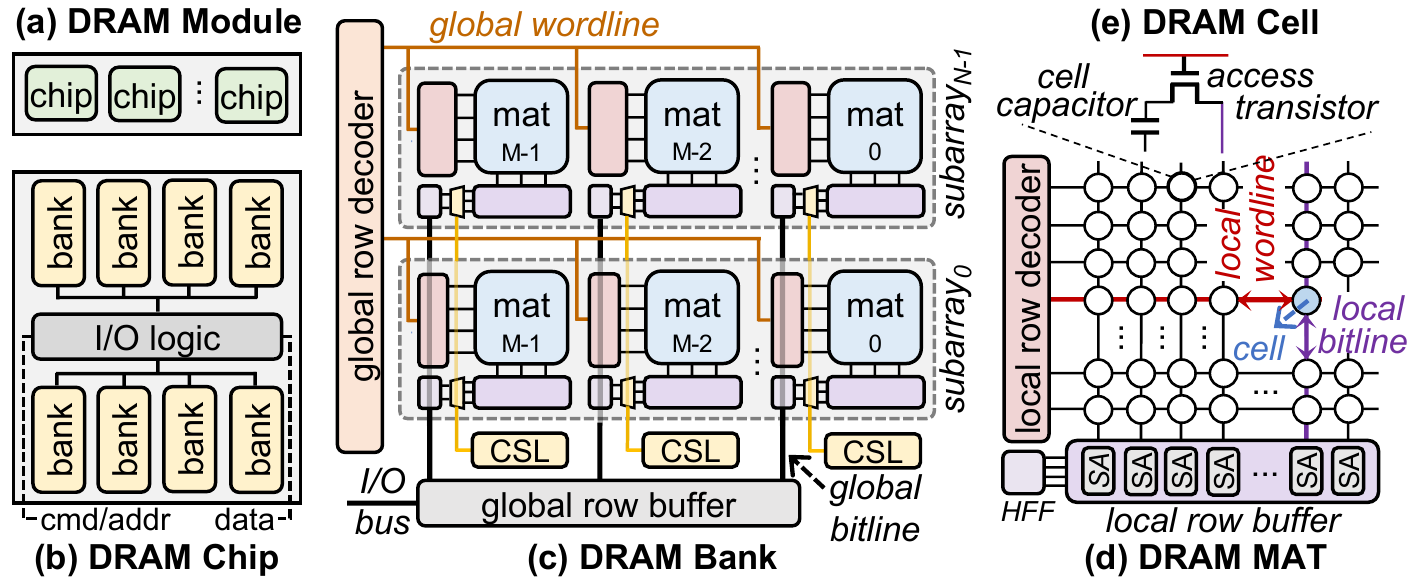}
    \caption{DRAM module, chip, and bank organization, as depicted in~\cite{oliveira2024mimdram}}
    \label{fig:dram_organization}
\end{figure}

A DRAM bank consists of a \emph{global \atbcr{1}{row buffer}} (\gf{or} \emph{prefetch buffer}), \atbcr{1}{a} \emph{global row decoder} (or a global wordline driver), and {multiple \emph{subarrays}}. \atbcr{1}{The global row decoder drives a \emph{global wordline} signal in every subarray}. \atbcr{1}{\emph{Global bitlines} connect the DRAM cells in each subarray to the global row buffer via column select logic (CSL).} {Each subarray contains a set of \emph{local \atbcr{1}{row decoders}} \atbcr{1}{(or local wordline drivers)}, a \emph{local row buffer}, which is an array of sense amplifiers (SAs), \emph{helper flip-flops (HFFs)}, and \emph{mats} (d). Inside every mat, \emph{DRAM cells} are placed in a two-dimensional array of \emph{local wordlines} and \emph{\atbcr{1}{local} bitlines}. A DRAM cell (e) \gf{is comprised of}
an \emph{access transistor} and a \emph{cell capacitor}. DRAM cells that lie on the same local wordline across different mats form a \emph{DRAM row} (not shown in the figure).}

\subsection{Accessing DRAM}
\label{sec:background:accessing}

\gf{The memory controller accesses DRAM in two steps.} First, the memory controller sends an $ACTIVATE$ ($ACT$) command with a \emph{row address} {to} \omcr{1}{open} (i.e., make accessible) a DRAM row. {The \atbcr{1}{global} row decoder drives the \atbcr{1}{global} wordline \atbcr{1}{(master wordline)} corresponding to the higher-order bits of the row address. The master wordline enables \emph{a single local wordline}, \atbcr{1}{addressed by the lower-order bits of the row address,} in every mat in a subarray. \new{\atbcr{1}{A driven local wordline} enables the access transistors \atbcr{1}{of all cells in the DRAM row}, causing the cells to share their charge with their bitlines and the sense amplifiers to read the values in the cells.} \revdel{Consequently, all cells on the same DRAM row start sharing the charge in their capacitors with their bitlines as the access transistors are enabled. Then, the local sense amplifiers amplify the voltage difference on the bitline and read out the value encoded in each cell}}{Second, the memory controller sends a $READ$ command with a \emph{column address} to retrieve multiple bytes of data (e.g., \SI{8}{\byte} for an x8 DDR4 chip~\cite{jedec2017ddr4}) from the \omcr{1}{open} DRAM row.} {The $READ$ command moves the data in the local \atbcr{1}{row buffer}, over the \atbcr{1}{helper flip-flops (HFFs)}, to the \atbcr{1}{global row buffer (the prefetch buffer)}. Thus, the \emph{throughput} of internal DRAM data transfers (i.e., between the global and local \atbcr{1}{row buffer}) is \new{constrained} by the number of HFFs per mat.\revdel{Once the data is in the prefetch buffer, the DRAM chip sends the data to the memory controller\revdel{over the memory channel in a \emph{burst}}.}}

\noindent
\textbf{Row buffer.} {Once a row is \omcr{1}{opened} (i.e., \atbcr{1}{the} row is buffered \atbcr{1}{in the local row buffer}), subsequent $READ$ and $WRITE$ commands targeting the row can be served at a fast rate. An access that targets an \omcr{1}{open} row is a \emph{row buffer hit} and an access that targets a row other than the \omcr{1}{open} row in a bank is a \emph{row buffer conflict}.} 

\noindent
\textbf{Accessing another row.}
When a bank already has an \omcr{2}{open} row, and the memory controller wants to access \tacorevcommon{another} row, the memory controller first issues a $PRECHARGE$ ($PRE$) command to \omcr{1}{close} the \omcr{1}{open} DRAM row.

\subsection{Data Transfer Bursts}
\label{sec:background:datatransfer}

DRAM modules transfer data on the memory channel over multiple \emph{interface \atbcr{1}{clock} cycles}. For example, a $READ$ command transfers a cache block (e.g., 512 bits) over eight interface cycles in DDR4~\cite{micron2014ddr4}. Each such transfer is referred to as a burst, and the \emph{burst length} defines the number of double-data-rate \tsub{(DDR)} interface cycles it takes to transfer the data. 
A cache block is \tsub{divided into equally-sized pieces and placed in different} chips (e.g., if there are eight chips, each chip receives $\frac{1}{8}$ of the cache block). These \tsub{equally sized} blocks are further split into multiple mats inside a bank in the chip. We depict how a cache block is scattered across multiple chips and mats for a DDR4 module in {\gf{Fig.}~\ref{fig:cache-block-placement} \gf{(left)}.
\gf{Fig.~\ref{fig:cache-block-placement} (right) shows the timing diagram of the command and data buses during a $WRITE$ transfer.}}

\begin{figure*}[!ht]
    \centering
    \includegraphics[width=0.8\linewidth]{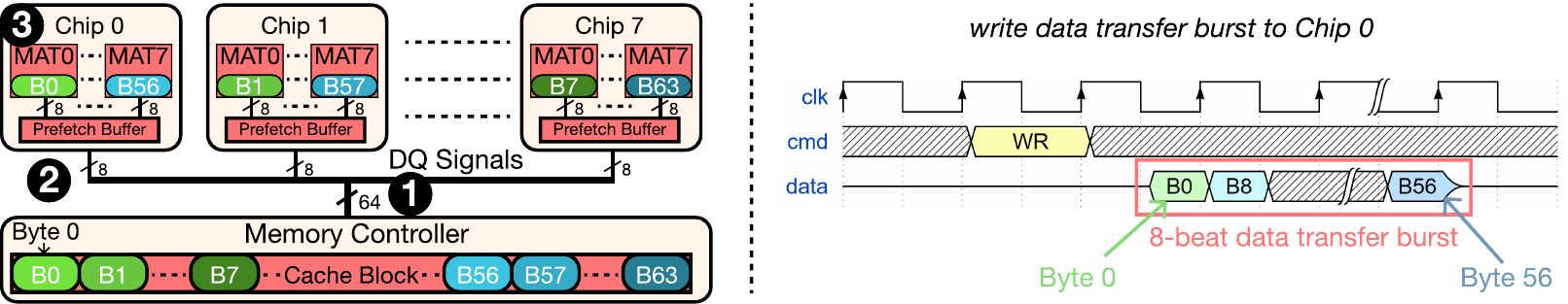}
    \caption{\omcr{1}{Example} cache block placement \atbcr{1}{in DRAM mats (left) and diagram depicting an 8-cycle data transfer burst to Chip 0 (right). ``B'' means ``byte''.}
    }
    \label{fig:cache-block-placement}
\end{figure*}

\gf{A DRAM data transfer happens in three steps (we use a $WRITE$ transfer as an example; a $READ$ data transfer happens \tsub{analogously}).} \gf{First,} the memory controller drives 64 \emph{DQ} \atbcr{1}{(data)} {signals} to transfer a 64-bit portion of the cache block in each \emph{beat} \atbcr{1}{(i.e., data transmitted in one DDR interface cycle)} of the \atbcr{1}{data transfer} burst\tsub{~\dingOne{}}. Second, each chip receives 8 bits in a beat\tsub{~\dingTwo{}}. A chip accumulates 64 bits during the burst in its prefetch buffer. Third, these 64 bits are copied into the mats inside the chip\tsub{~\dingThree{}}. \tsub{In our example, o}nly 8 bits are transferred in a burst to (from) a mat with a $WRITE$ ($READ$) command. Thus, the \emph{maximum DRAM throughput} can \emph{only} be obtained if every mat contributes 8 bits to the data transfer burst.

\subsection{The \tFAW{} Timing Parameter}
\label{sec:background:tfaw}

\new{DDRx specifications} \atbcr{1}{(e.g., DDR4~\cite{jedec2017ddr4} and DDR5~\cite{jedec2020ddr5})} define the \tFAW{} timing parameter, which specifies the time window where no more than four $ACT$ commands are allowed (i.e., the memory controller can only schedule four $ACT$ commands in any \tFAW{}-wide time window). \new{\tFAW{} allows \atbcr{1}{a DRAM chip to} correctly provide the chip's various components with the power required to activate large DRAM rows. \tFAW{} \atbcr{1}{typically} limits row activation frequency and diminishes memory-level parallelism, degrading the performance of memory-intensive workloads~\cite{kaseridis2011minimalist}.}

\section{Motivation}
\label{sec:motivation}

\revdel{\gf{\new{A} DRAM system exploit\new{s} applications' spatial locality on two levels. First, the system exploits \emph{word-level spatial locality} by transferring an entire cache block, with multiple words\footnote{\new{We refer to each non-overlapping 64-bit portion of data in a cache block as a word.}}, from DRAM to the memory controller over a single memory burst (\emph{coarse-grained DRAM access}). Second, the system exploits \emph{cache block-level spatial locality} by activating an entire DRAM row, with multiple cache blocks, during a read/write operation (\emph{coarse-grained DRAM activation}). \revdel{Coarse-grained DRAM access and activation can increase DRAM throughput and reduce main memory access latency for applications with enough spatial locality. However, it can also increase energy consumption for applications with inherent low spatial locality~\cite{rhu2013locality,yoon2011adaptive,zhang2014half,yoon2012dynamic,kumar2012amoeba}.}}}

\gf{We study the impact of coarse-grained DRAM \atbcr{1}{data transfer} (\emph{Coarse-DRAM-Transfer}) and activation (\emph{Coarse-DRAM-Act}) in 41 single-core workloads from a \new{variety} of \revdel{workload }domains\revdel{ from the DAMOV~\cite{oliveira2021damov}, SPEC2006~\cite{spec2006}, and SPEC2017~\cite{spec2017} benchmark suites} (see~\cref{sec:methodology} for our\revdel{ evaluation} methodology). We compare their energy consumption to a system that performs
(i) fine-grained DRAM \atbcr{1}{data transfer} (\emph{Fine-DRAM-Transfer}) \omcr{1}{at} word granularity\revdel{(i.e., the memory controller can transfer individual words from/to DRAM)}, and 
(ii) fine-grained DRAM activation \nis{(}\emph{Fine-DRAM-Act}\nis{)} \omcr{1}{at} mat granularity\revdel{ (i.e., \new{DRAM mats can be activated individually})}.}


\revdel{We conclude that current systems that perform coarse-grained DRAM accesses significantly waste energy by transferring a fixed number of words from DRAM to the processor, which are not entirely \new{used by the processor}.}

\gf{Fig.~\ref{fig:cache-block-word-usage} (left) shows the DRAM access energy across workloads from our three benchmark suites for the \emph{Coarse-DRAM-Transfer} system, normalized to the \emph{Fine-DRAM-Transfer} system. We observe that the DRAM access energy of the \emph{Coarse-DRAM-Access} system is 1.27$\times$ that of the  \emph{Fine-DRAM-Access}. The large increase in energy consumption in the \emph{Coarse-DRAM-Access} system is caused by retrieving words in a cache block that the processor does \emph{not} entirely use. This leads to a 45\% increase in the data movement between DRAM and the CPU in the \emph{Coarse-DRAM-Access} system, on average. We conclude that current systems that perform coarse-grained DRAM accesses and activation \emph{significantly} waste energy by transferring and activating a fixed number of words from DRAM to the processor, which are not entirely \new{used by the processor}.}

\begin{figure}[!ht]
    \centering
    \includegraphics[width=\linewidth]{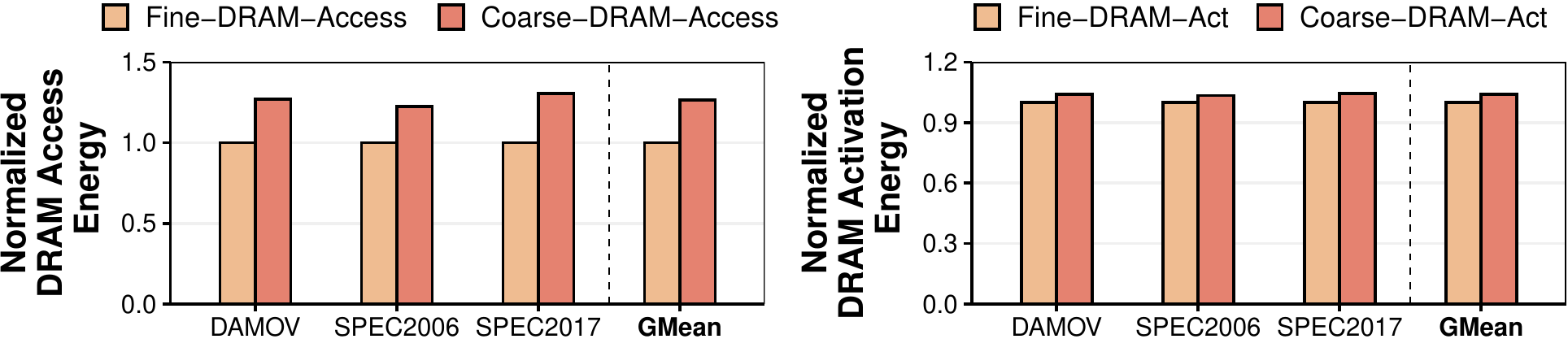}
    \caption{\gf{Normalized DRAM access (left) and DRAM activation (right) energy consumption.}}
    \label{fig:cache-block-word-usage}
\end{figure}

\gf{Fig.~\ref{fig:cache-block-word-usage} (right) shows the DRAM activation energy across workloads from our three benchmark suites for the \emph{Coarse-DRAM-Act} system, normalized to the \emph{Fine-DRAM-Act} system. We observe that the DRAM activation energy of the \emph{Coarse-DRAM-Act} system is 1.04$\times$ that of the \emph{Fine-DRAM-Act} system. Like the system that performs coarse-grained DRAM accesses, the increase in energy consumption in the coarse-grained DRAM activation system is caused by activating a large\new{,} fixed-size DRAM row that the processor does \emph{not} entirely \new{use}. As prior works~\cite{lee2017partial,ghose2019demystifying,mutlu2013memory,mutlu2007memory,mutlu2007stall,sudan2010micro,yuan2009complexity,nesbit2006fair,subramanian2016bliss} show, such an increase in energy consumption when executing coarse-grained DRAM activation is because modern memory-intensive workloads with irregular access patterns suffer from low spatial locality, which reduces the benefit of a large DRAM row buffer. We conclude that systems that employ coarse-grained DRAM activation \new{suffer} from energy inefficiency.}

\begin{figure*}[!b]
    \centering
    \includegraphics[width=0.9\linewidth]{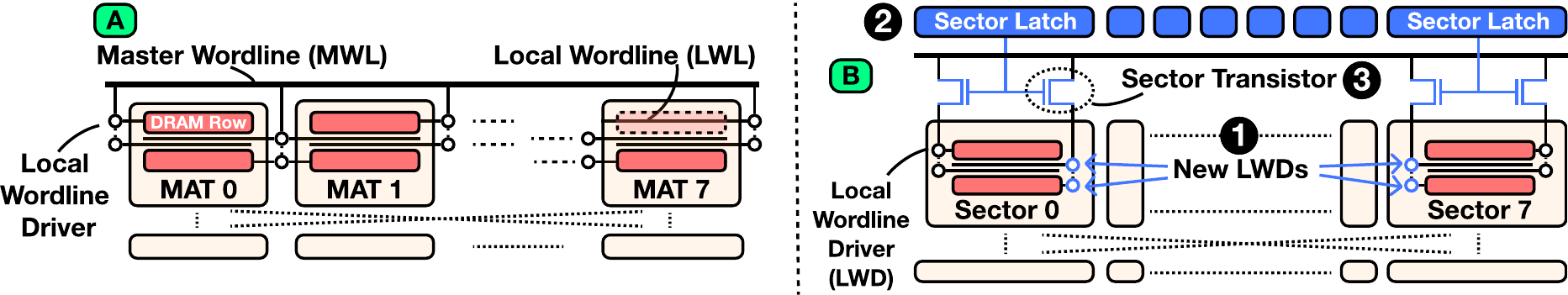}
    \caption{Wordline organization in a \atbcr{1}{conventional DRAM} subarray \atbcr{1}{(left), in a \X{} subarray (right)}}
    \label{fig:sectored-activation-merged}
\end{figure*}

\subsection{\gf{Enabling Fine-Grained DRAM: Challenges and Limitations}}

\label{sec:motivation:limitations}

\gf{Efficiently enabling fine-grained DRAM \atbcr{1}{data transfer} and activation can significantly improve system energy. However, to do so, we must overcome three main challenges.}

\gf{\textbf{(1) Maintaining high DRAM throughput.} Current DRAM systems leverage coarse-grained \atbcr{1}{data transfers} to maximize DRAM's throughput\revdel{ since throughput directly impacts applications' performance. Thus, it is critical to sustain high DRAM throughput when enabling fine-grained DRAM}. Enabling fine-grained DRAM \atbcr{1}{in a straightforward way, e.g., by placing a piece of the cache block stored by a DRAM chip in the same mat instead of distributing the piece across multiple mats,} reduces DRAM throughput as one mat contributes only 
a fraction of the total DRAM internal throughput (\cref{sec:background:datatransfer}).
\new{This} issue can be alleviated by increasing the number of \new{the} HFFs. However, this approach is \omcr{2}{costly} since it severely complicates DRAM array routing~\cite{zhang2014half,lee2017partial,vogelsang2010understanding}.}

\gf{\textbf{(2) Incurring low DRAM area overhead.} DRAM manufacturing is highly optimized for density and cost~\cite{li.micro17,mutlu2013memory,mandelman2002challenges}. While enabling fine-grained DRAM, one must avoid applying intrusive modifications to the DRAM array since such modifications are \new{difficult to integrate into} real designs.}

\gf{\textbf{(3) Fully exploiting Fine-Grained DRAM.} The energy waste of coarse-grained DRAM systems stems from rigid DRAM \atbcr{1}{data transfer} and activation granularities. Thus, a fine-grained DRAM system must enable flexible DRAM \atbcr{1}{data transfer} \emph{and} activation granularities for \emph{both} read \new{and} write operations to eliminate such energy waste. However, \new{integrating fine-grained DRAM \gf{into current} system\gf{s}} is challenging as \new{systems are typically} designed to access DRAM \omcr{1}{at} cache block granularity.}

\gf{\omcr{1}{P}rior works~\cite{lee2017partial,zhang2014half,zhang2017enabling,alawneh2021dynamic,son2014microbank,oconnor2017fine,chatterjee2017architecting,cooper2010fine,ha2016improving,udipi2010rethinking} propose different mechanisms to enable fine-grained DRAM substrate\new{s}, aiming to alleviate the energy waste caused by coarse-grained DRAM. Such works can be divide\nis{d} into two broader groups: 
(1) works that propose \emph{intrusive} modifications to the DRAM array circuit and organization \new{(e.g., new DRAM interconnects, considerably more HFFs)}~\cite{zhang2017enabling,alawneh2021dynamic,son2014microbank,oconnor2017fine,chatterjee2017architecting} and 
(2) works that aim to enable coarse-grained DRAM \emph{without intrusive} modifications to DRAM~\cite{cooper2010fine,udipi2010rethinking,zhang2014half,ha2016improving, lee2017partial}. The intrusive DRAM \new{modifications} proposed by the first group lead to significant DRAM area overheads, which \new{makes it difficult to integrate \omcr{1}{the first} group of works into real DRAM designs}.}

\gf{Table~\ref{table:comparison} qualitative\new{ly} compares how prior works from the second group address the three challenges of enabling fine-grained DRAM. We observe that no prior work can \emph{simultaneously} provide 
(i) high DRAM throughput (FGA~\cite{cooper2010fine} and SBA~\cite{udipi2010rethinking} change the cache block mapping such that DRAM transfers can be served from only one mat, but \new{reduce the throughput of data transfers by doing so});
(ii) low area overhead (HalfDRAM~\cite{zhang2014half} and HalfPage~\cite{ha2016improving} require changes to the number and organization of DRAM's HFFs, leading to non-negligible area overheads); and
(iii) mechanisms that fully exploit fine-grained DRAM (PRA~\cite{lee2017partial} only enables fine-grained DRAM data transfer and activation for write operations; HalfDRAM, HalfPage, FGA, and SBA still impose a rigid DRAM \new{data transfer} granularity). We conclude that no prior work efficiently enables fine-grained DRAM \atbcr{1}{access (i.e., both data transfer and activation)}.}

\begin{table}[ht]
    \tempcommand{0.4}
    \centering
    \caption{\X{} vs.\ prior works}
    \label{table:comparison}
    \footnotesize
 \resizebox{1\linewidth}{!}{
\begin{tabular}{@{}lcccc@{}}
\toprule
\textbf{}          & \textbf{\begin{tabular}[c]{@{}c@{}}High \atbcr{1}{(100\%)}\\Throughput\end{tabular}} & \textbf{\begin{tabular}[c]{@{}c@{}}Low \atbcr{1}{(<2\%)}\\Area Overhead \end{tabular}} & \multicolumn{2}{c}{\textbf{\begin{tabular}[c]{@{}c@{}}Fully Exploit\\  Fine-Grained DRAM\\\end{tabular}}} \\
& & & \omcr{2}{Activation} & \omcr{2}{Data Transfer} \\
\midrule
FGA~\cite{cooper2010fine}& \XSolidBrush& \Checkmark& \Checkmark&\XSolidBrush\\
SBA~\cite{udipi2010rethinking}& \XSolidBrush& \Checkmark& \Checkmark&\XSolidBrush\\
HalfDRAM~\cite{zhang2014half}   & \Checkmark& \XSolidBrush& \Checkmark&\XSolidBrush\\
HalfPage~\cite{ha2016improving}   & \Checkmark& \XSolidBrush&\Checkmark& \XSolidBrush\\
PRA~\cite{lee2017partial}        & \Checkmark& \Checkmark& \Checkmark&\XSolidBrush\\ \midrule
\textbf{This Work} & \CheckmarkBold& \CheckmarkBold& \CheckmarkBold & \CheckmarkBold\\ \bottomrule
\end{tabular}
}
\end{table}

\gf{Our \textbf{goal} is to address prior works' limitations while efficiently mitigating the energy consumed by transferring unused data on the memory \new{channel} and activating unused DRAM cells.\revdel{ We aim to maintain high DRAM throughput, incur low area overhead, and fully exploit fine-grained DRAM access and activation.} To this end, we develop \X{}, a new, practical, and high-performance fine-grained DRAM substrate.}

\section{\X{}}

\revdel{We introduce \X{}, the first fine-grained DRAM substrate that mitigates the energy inefficiencies of coarse-grained DRAM while improving system performance.} \unsure{Is this in the Intro?}\new{We leverage two key observations regarding DRAM chip design \new{to implement \X{} at low cost}. First, we observe that\revdel{the hierarchical structure within a DRAM subarray} DRAM mats naturally split DRAM rows into fixed-size portions. Second, the DRAM I/O circuitry already implements a mechanism to select one portion of a cache block to transfer it in one beat \atbcr{1}{(i.e., data transmitted in one DDR interface cycle)} of a burst.}

\revdel{We leverage these two observations to implement \X{}\revdel{, a fine-grained DRAM substrate that enables energy savings and performance improvements via fine-grained DRAM activation and access} at \gf{a} low \gf{area} overhead. }\gf{\X{} consists of two new mechanisms implemented in a DRAM chip:}
{\XMechOne{} (\XMO{}) and \XMechTwo{} (\XMT{}). \XMO{} enables fine-grained control over the activation of sectors in DRAM by making minimal modifications to how local wordlines are driven. \XMT{} enables fine-grained
control over data transfer bursts by transferring only the portions of a cache block that correspond to the activated sectors in the DRAM chip.} 

\new{We expose the two mechanisms to the memory controller such that the system can benefit from \X{}, with \emph{no} changes to the physical DRAM interface and only small changes to \gf{DRAM} interface specification.}

\subsection{\underline{S}ectored \underline{A}ctivation
}
\label{sec:sectored-activation-mechanism}

\new{\gf{Fig.}~\ref{fig:sectored-activation-merged}-A depicts the architecture of a DRAM array with 8 mats in one subarray~\cite{koo2012ddr4sdram,ha2016improving,keeth2008dram,zhang2014half,oconnor2017fine}. We make a \textbf{key observation}: \revdel{The hierarchical structure within the subarray naturally} DRAM mats split DRAM rows into fixed-size portions. \XMO{} augments these portions by allowing them to be activated individually. We refer to these augmented portions as \emph{sectors}.}


\noindent
\textbf{\textbf{$\textbf{SA}$} Design.} To implement \XMO{}, we propose minor modifications to \new{the existing} architecture. \new{Fig.~\ref{fig:sectored-activation-merged}-B depicts our modifications to the DRAM subarray with 8 mats (the modifications are highlighted in blue color).\footnote{\new{We consider DRAM subarrays with 8 sectors. See~\cref{sec:discussion-finer-granularity-sectors} for a discussion \omcr{1}{of} a finer-granularity activation mechanism.}}} \new{First, we insert new local wordline drivers (LWDs) \atbcr{1}{(\dingOne{})} such that each LWD drives only one local wordline (LWL). Thus,\revdel{by enabling a single LWD, we make sure that only the cells that belong in a single mat get activated} when a single LWD is enabled, only the cells in a single sector are activated as opposed to cells from \omcr{1}{all mats in a subarray} getting activated in the existing architecture (e.g., LWDs between mat 0 and mat 1 in Fig.~\ref{fig:sectored-activation-merged}-A drive two LWLs that extend onto both of the mats).} \unsure{What is the sector latch latching?}Second, \atbcr{1}{to select which sectors are opened with an $ACT$ command,} we place one \emph{sector latch} (\dingTwo{}) for every sector in the horizontal direction. Third, we isolate the master wordline (MWL) from the LWDs \nis{using \emph{sector transistors}} (\dingThree{})\atbcr{1}{, such that a driven MWL does \emph{not} enable all LWDs in all sectors, but only enables LWDs in the open sector(s)}. \new{With these \tacorevd{three modifications, a sector with a set (i.e., logic-1) sector latch is}}\changev{\ref{q:r4q2}} activated when the MWL is driven (with an $ACT$ command) because two sector transistors connect the MWL to the LWDs.

\noindent
\textbf{Exposing $\textbf{SA}${} \atbcr{1}{to the memory controller}.} {To make use of \XMO{}, the memory controller needs to \new{control} sector latches.\revdel{We propose exposing the control of sector latches to the memory controller via the unused signals in DDR4 command encoding~\cite{jedec2017ddr4}.} \new{To implement \XMO{} with no modifications to the DRAM interface signals, we use the unused bits in the $PRE$ command's encoding~\cite{jedec2017ddr4} to encode the \emph{sector bits}. Each sector bit \omcr{1}{identifies/encodes if a sector latch is set or reset.} \atbcr{1}{T}he memory controller sends \atbcr{1}{a bitvector of sector bits with every} $PRE$ \atbcr{1}{command} to \atbcr{1}{the \X{} chip. These sector bits are used for the $ACT$ command that follows the $PRE$ command.} 
When the bank is closed (i.e., \atbcr{1}{there are} no \omcr{1}{open} rows in the bank), the memory controller schedules a $PRE$ command before the first $ACT$ command to convey the sector bits \atbcr{1}{to the DRAM chip}. The \atbcr{1}{minimum timing delay} between successive $PRE$ and $ACT$ commands targeting the same bank (\atbcr{1}{dictated by} \tRP, $\sim{}\!13ns$~\cite{jedec2017ddr4,micron2014ddr4}) \atbcr{1}{is sufficiently long\footnote{\atbcr{1}{We determine \tRP{} to be sufficiently long based on the overall latency of a READ command in a conventional DRAM chip. A READ command 1) propagates from the memory controller to the DRAM chip, 2) accesses data in a portion of the row buffer in the corresponding bank, and 3) sends the data back to the memory controller. In the DDR4 standard~\cite{jedec2017ddr4}, the latency between issuing a READ and the first data beat appearing on the data bus is defined as $tAA$ (\SI{12.5}{\nano\second}). Because sector bits need \emph{only} to propagate from the memory controller to a DRAM bank (and not back to the memory controller from a bank), the \tRP{} timing parameter is likely longer than what is needed (which we estimate as half $tAA$ or \SI{6.25}{\nano\second}) for sector bits to propagate from the memory controller to a DRAM bank.}} such} that the sector bits can propagate from the DRAM chip's inputs to the sector latches \atbcr{1}{before the $ACT$ command following a $PRE$ command is issued to the same DRAM bank}.}} \Copy{R4/6}{\changev{\ref{q:r4q6}}\tacorevd{To issue regular, row-level DRAM commands (e.g., a periodic refresh or an activate command), the memory controller simply sets all sector bits (to enable all sectors) before issuing the row-level command.}}

\feedback{I think we need to bank more in this and extend this as a subsubsection by itself. THis is the primary sourc eof perf. benefits of sec.dram.}

Because activating one sector requires considerably less power \omcr{1}{consumption} (\cref{sec:results-dram-power}) than activating all sectors, the \tFAW{} timing constraint can be relaxed \tsub{to allow for, within a \tFAW{}, a larger number of $ACT$ commands that activate fewer than all eight sectors in a DRAM row~\cite{zhang2014half,lee2017partial,oconnor2017fine,mathew2017using,mathew2020using}. \atbcr{1}{Section~\ref{sec:methodology:simulation} describes how exactly we relax \tFAW{} based on how many sectors an $ACT$ command activates.}}

\subsection{\underline{V}ariable \underline{B}urst \underline{L}ength}
\atbcr{2}{\XMT{}'s} \atbcr{1}{goal is to allow a} DRAM chip to transmit (i.e., $READ$) and receive (i.e., $WRITE$) data in variable length bursts\footnote{\atbcr{1}{Commodity DDR4 chips already implement a relatively constrained version of \XMT{} called \emph{burst-chop}. Burst-chop enables 256-bit (in 4-beat bursts) data transfers~\cite{jedec2017ddr4}.}} such that each beat of the burst transfers only the data corresponding to one of the enabled sectors. 

\noindent
\textbf{$\textbf{VBL}$ design.} \gf{Fig.}~\ref{fig:variable-burst-length} depicts the I/O read/write circuitry \gf{of} a modern DRAM chip~\cite{micron2014ddr4}. In such a chip, data is first \gf{moved} from the DRAM array to the \emph{Read FIFO} (\dingOne{}) \atbcr{1}{with every} \nis{$READ$ \atbcr{1}{command}}. The \emph{Read FIFO} comprises eight entries, and each entry stores the data that will be transmitted over the DQ pins in one beat of the data transfer burst. The \emph{Read MUX} (\dingTwo{}) selects one entry in the \emph{Read FIFO} based on the value of the \emph{burst counter} (not shown in the figure), which counts the number of beats in the transfer\revdel{ (e.g., from 0 to 7)}. 

\begin{figure}[ht]
    \centering
    \includegraphics[width=1.0\linewidth]{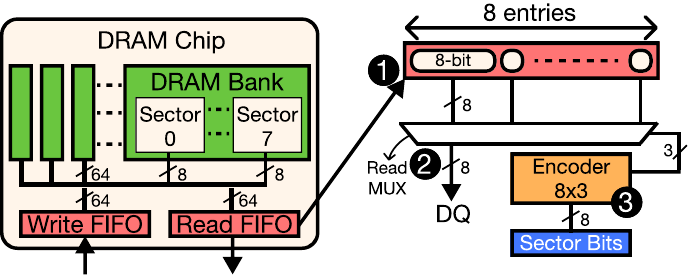}
    \caption{\atbcr{1}{I/O circuitry of a DRAM chip and} \XMechTwo{}}
    \label{fig:variable-burst-length}
\end{figure}

\gf{By studying the I/O read/write circuitry \gf{of} modern DRAM chips, we observe that \atbcr{1}{a} DRAM chip selects (using the burst counter) individual entries in the \emph{Read FIFO} to drive the DQ pins within a beat~\cite{micron2014ddr4}. Based on \atbcr{1}{this observation},} \XMT{}'s \textbf{key idea} is to \atbcr{1}{slightly} modify the \atbcr{1}{DRAM chip's Read FIFO entry selection criteria}. We replace the burst counter with an 8$\times$3 encoder (\dingThree{}) that takes sector bits as input, and outputs only the indices for the \emph{Read FIFO} entries that contain data from one of the \omcr{1}{closed} sectors. \atbcr{1}{Using the encoder}, the \emph{Read MUX} skips the entries in the \emph{Read FIFO} that correspond to the \omcr{1}{closed} sectors, driving the DQ pins \emph{only} with the data that comes from the \omcr{1}{open} sectors (\dingFour{} in Fig.~\ref{fig:variable-burst-length}, right). Since the \emph{Write FIFO} is organized in the same way as the \emph{Read FIFO}, \atbcr{1}{\XMT{}} reuses the same encoder for $WRITE$ transfers to correctly fill the entries in the \emph{Write FIFO}.

\noindent
\textbf{Exposing $\textbf{VBL}$ \atbcr{1}{to the memory controller}.} To use \XMT{}, the memory controller and the DRAM chip need to agree on the burst length \emph{before} the data transfer starts. This is important for both parties to calibrate their I/O drivers correctly and capture the signals on the high-frequency DRAM interface~\cite{jedec2017ddr4, jedec2020ddr5, jedec2007ddr3}.
We use the sector bits that are already communicated to the DRAM chip by \gf{fine-grained DRAM activation} operations to determine the burst length (\cref{sec:sectored-activation-mechanism}) \new{of data transfers, before the transfers start}.

We make two modifications. {First, we implement a low overhead 8-bit \emph{popcount} circuit~\cite{frontini2018very,dalalah2006new} in both the DRAM chip and the memory controller to count the number of set sector bits in the DRAM bank targeted by a $READ$ or a $WRITE$. The popcount circuitry requires \emph{only} 34 logic gates to \gf{be} implement\gf{ed}~\cite{dalalah2006new}, introduc\gf{ing} \omcr{1}{almost} negligible area overhead. Second, we extend the \emph{bank state table} of the memory controller with sector bits. The bank state table already contains metadata, such as the address of the enabled row, for every bank. The additional storage requirement for sector bits in the bank state table is small, only 128 bits (8 bits for each \omcr{1}{of the 16} banks~\cite{micron2014ddr4,jedec2017ddr4}).}

\section{System Integration}
\label{sec:system-integration}
{We describe the challenges in integrating \X{} into a typical system and propose solutions. We assume that the system uses a DDR4 module with \emph{eight chips} as main memory and that each chip has \emph{eight sectors} to explain the challenges and our solutions clearly.{\footnote{\tsub{In \cref{sec:discussion}, we discuss how \X{} can be integrated into systems with different parameters (e.g., more sectors per chip).}}} Since there are eight sectors in every chip, {one sector from each DRAM chip \unsure{together with what?}\new{collectively}} store\tsub{s} \emph{one word} (64 bits) of the cache block.}



\noindent
\new{\textbf{Integration challenges}.} We identify two challenges in integrating \X{} \gf{into a }system. First \atbcr{1}{(\cref{sec:tracking-valid-words})}, 
{\atbcr{1}{to benefit from \X{}'s potential energy savings, the system and main memory (DRAM) must conduct data transfers at sub-cache-block granularity (e.g., transfer one or multiple words).}
Therefore, a cache block may have both \emph{valid} (up-to-date) and \emph{invalid} (stale or evicted) words present in \atbcr{1}{system caches}. However, \atbcr{1}{caches} keep track of the valid on-chip data at cache block granularity. This granularity is too coarse to keep track of valid words in a cache block.} Second \atbcr{1}{(\cref{sec:accurate-word-retrieval})}, {because some words in a cache block can be invalid, references to these words (e.g., made by load/store instructions) would result in a cache miss. This can \revdel{increase the average memory latency and }induce performance overheads. \revdel{Thus, the system must carefully retrieve all words in a cache block that will be referenced by load/store instructions during the cache block's residency in the caches (i.e., from the moment the cache block is brought to on-chip caches until the block is evicted).}}

{We propose the following minor \atbcr{1}{system} modifications to overcome \X{}'s integration challenges. First, to track which words in a cache block are valid, we extend cache blocks with additional bits \atbcr{1}{each of which indicates} if a 64-bit word in the cache block is valid. Second, to accurately retrieve all \emph{useful} words in a cache block, i.e., words that will be used \new{until the cache block is evicted}, \new{we develop \atbcr{1}{two techniques}: (i) \nis{\atbcr{1}{Load-Store Queue (LSQ)} Lookahead, and (ii) Sector Predictor.}}

\subsection{Tracking Valid Words \atbcr{1}{in the Processor}}
\label{sec:tracking-valid-words}
{Since \atbcr{1}{a} \X{}\atbcr{1}{-based system} can retrieve individual words of a cache block from DRAM, \atbcr{1}{system} caches must store data at a granularity that is finer than the typical 512-bit granularity. One straightforward approach to allow \atbcr{1}{finer-}granularity storage in caches is to reduce the cache block size from 512 bits to \atbcr{1}{the size of a word (e.g., 64 bits)}. However, \atbcr{1}{for the same cache size, doing so} requires implementing $8\times{}\!$ as much storage for \emph{cache block tags}, which introduces significant area overhead. Instead, we extend cache blocks with just eight additional bits \atbcr{1}{each of which} indicates whether a word in the cache block is valid or invalid, using sectored caches~\cite{liptay1968structural, smith1987line, alpert1988performance, hill1984experimental, rothman2000sector,kadiyala1995dynamic,anderson1995two,rothman2002minerva,kuangchih1997on}.}

\atbcrcomment{There used to be a figure here. I thought it was not necessary as we are describing very basic things.}
\noindent
\atbcr{1}{\textbf{Sector cache organization.}} We \atbcr{1}{add storage for} {eight sector bits} \atbcr{1}{to all system caches}. {We store sector bits in a CAM array (similar to how cache block tags are stored). This allows \atbcr{1}{a cache} to find out if a \emph{sector is missing} (i.e., a word is invalid) in \atbcr{1}{a} cache block in parallel with the tag lookup.} We divide a cache block into eight sectors, each corresponding to a word. {Thus, we only add eight sector bits to \atbcr{1}{a} cache block, \atbcr{1}{which incurs low storage overhead} (\cref{sec:results-area}).}

\noindent
\atbcr{1}{\textbf{Sector cache operation.}}
\atbcr{1}{We describe the three-step process performed by a memory request to access a word in the highest-level sector cache (i.e., the L1 cache).} {First, the processor sends a memory request with \atbcr{1}{a memory address} and a vector of eight sector bits \atbcr{1}{to the highest-level cache}. \new{The sector bits \atbcr{1}{identify} the} word\new{s} in the cache block that the \atbcr{1}{processor core} demands.} Second, the \atbcr{1}{L1 cache uses  the memory address to identify the addressed cache set}. Third, the L1 cache \atbcr{1}{uses the cache block tag component of the memory address and the sector bits to} \atbcr{1}{access the words requested by the processor}. \atbcr{1}{The third step can result in three different scenarios: 1) if both the tag and the sector bits match one of the cache blocks in the cache set (i.e., there is both a tag and a sector bit match), the cache has the word that the processor core demands and this is a \emph{sector hit}, 2) if there is a tag match but no sector bit match, the cache has to request the missing sectors from a lower-level cache or main memory and this is a \emph{sector miss}, and 3) if there is no tag match, this is a \emph{cache miss}.} 

\noindent
\atbcr{1}{\textbf{Sector misses.} On a sector miss, the cache controller creates a memory request to retrieve the missing sector(s) from a lower-level cache or main memory. The cache controller determines the missing sector(s) by bitwise AND'ing the memory request's (the request that triggers the sector miss) sector bits and the sector bits that are \emph{not} set in the cache block. When the created memory request returns from a lower-level cache, the cache controller sets the cache block's missing sector bits.}

\noindent
\new{\textbf{Sector cache compatibility.} Sector caches do not require any modifications to existing \emph{cache coherence protocols} \tsub{(we explain how in the next paragraph)}. Sector caches are compatible with existing SRAM ECC schemes~\cite{mano1983submicron,tendler2002power4,moore1986review}, as the invalid words (i.e., missing sectors) in a cache block can still be used to correctly produce a codeword.}

\noindent
\changev{\ref{q:r2q1}}\Copy{R2/1}{\tacorevb{\tsub{\microrev{\textbf{Cache Coherence.} \X{} requires no modifications to existing cache coherence protocols that operate at the granularity of a cache block since cache coherence in \X{} is \atbcr{1}{still} maintained at cache block \atbcr{1}{granularity}. A processor core can only modify a sector in a cache block if the core owns the \omcr{1}{entire} cache block (e.g., the cache block is in the M state in a MESI protocol). A cache block shared across multiple cores may have different valid sectors among its copies in different private caches. However, this does not violate cache coherence protocols.}}}}

\revdel{
\new{We discuss various aspects (e.g., cache block management, coherency) of \X{}'s implementation of sectored caches in detail:}

\noindent
\microrev{\textbf{Cache Block Evictions.} When a cache block with a dirty sector is evicted from \new{a higher level} cache, the dirty sector in that cache block overwrites the copy of the same cache block in the \new{lower-level} cache and updates the sector bits of the copy of the cache block in the \new{lower level} cache.} \microrev{A cache block without any dirty sectors is simply invalidated as nothing needs to be written back to a lower level component (e.g., L2 cache) in the memory hierarchy.}

\noindent
\microrev{\textbf{Cache Inclusion Policy.} We implement \X{} using an inclusive caching policy for cache blocks. If a cache block is valid in a lower level cache, it is also valid in the higher level caches. A cache block might be invalid in a lower level cache, but it can be valid in a higher level cache. If a sector in a clean cache block is valid in a higher-level cache (e.g., L1), it is also valid in a lower level cache (e.g., L2). However, the CPU may update an invalid sector of a valid cache block in the higher level cache by executing store instructions (making the sector dirty in doing so). This dirty sector can be valid in the higher level cache and invalid in the lower level cache.}

\noindent
\microrev{\textbf{Cache coherence.} \X{} requires no modifications to existing cache coherence protocols that operate at the granularity of a cache block since cache coherence in \X{} is maintained at the granularity of a cache block. A core can only modify a sector in a cache block if the core owns the cache block (e.g., the cache block is in the M state in a MESI protocol). A cache block shared across multiple cores may have different valid sectors among its copies in different private caches. However, this does not violate cache coherence protocols.}

\noindent
\microrev{\textbf{Compatibility with SRAM ECC.} Sector caches are compatible with existing ECC schemes. The only difference between a sector cache’s cache block and a baseline cache’s cache block is that some portions (sectors/words) of the data in the sector cache’s cache block might be invalid.}

\microrev{We describe one way to support ECC in sectored caches. First, the cache controller fills the invalid sectors in a cache block with an all zeros or an all ones data pattern. Second, the cache controller performs an ECC encoding operation on the cache block’s data. Third, the cache controller updates the cache’s data array with the cache block’s data and parity bits (i.e., the codeword). To update an invalid sector in the cache block (e.g., when a LOAD request causes a sector miss and the missing sector is brought from a lower level cache), the cache controller i) reads the cache block from the data array, ii) updates the missing sector in the cache block, and iii) performs ECC encoding on the updated cache block.}

}
\noindent
\tsub{\microrev{\textbf{Other cache architectures.} There are numerous other multi-granularity cache architectures \cite{qureshi2007line,pujara2006increasing,rothman1999pool,seznec1994decoupled,kumar2012amoeba,inoue1999dynamically} that could be used instead of sectored caches in \X{} to improve cache utilization (e.g., by reducing the number of invalid words stored in a cache block) at the cost of increased storage for tags and hardware complexity~\cite{kumar2012amoeba}. We use sectored caches to minimize the storage and hardware complexity overheads in \X{} and leave the exploration of other cache architectures in Sectored DRAM to future work.}}

\subsection{Accurate Word Retrieval \atbcr{1}{from Main Memory}}
\label{sec:accurate-word-retrieval}

\new{With \atbcr{1}{sector caches}, a \atbcr{1}{\X{}-based system} can \atbcr{1}{transfer data at word granularity} \atbcr{1}{between components in the memory hierarchy (e.g., between the L1 and the L2 cache)} instead of \atbcr{1}{transferring data at} cache block granularity\revdel{ by executing load/store instructions}.\revdel{ \new{We describe the simplest way that the system can make use of \X{}.} 
\nis{To do so, the processor uses the sector bits \new{field in a} memory request and sets only the \new{bit (among all eight bits) that corresponds to the word a load/store instruction} needs to access. When the memory request is forwarded to the memory controller (by missing in all caches), the memory controller only retrieves the word indicated by the sector bits, benefiting from {the performance improvements provided by \XMO{} and energy savings provided by both \XMO{} and \XMT{}}.}
However, even though cache blocks are not fully utilized by the processor (\cref{sec:motivation}), workloads often access multiple words in a cache block during {the cache block's cache residency.}
\new{Therefore, retrieving only a single word with each memory access} can result in a high number of sector misses per cache block (due to load/store memory requests to the missing words in a cache block), causing multiple high latency DRAM accesses to retrieve the \new{missing words}.} However, retrieving cache blocks word-by-word from DRAM can reduce system performance compared to bringing cache blocks as a whole \gf{because the processor needs to complete multiple high latency DRAM accesses to retrieve a word (on a sector miss) as opposed to completing a single memory access to retrieve the whole cache block\revdel{ and supplying data for the rest of the requests from the caches}}.} 
\new{To minimize the performance overheads induced by \atbcr{1}{the additional} DRAM accesses and to better benefit from the energy savings provided by \X{}, we propose two mechanisms that greatly reduce the number of sector misses.}

\noindent
\new{\textbf{Load/Store Queue (LSQ) Lookahead.}}
The key idea behind LSQ Lookahead is to exploit the spatial locality in subsequent load/store instructions that target the same cache block. \atbcr{1}{A load or a store instruction typically \emph{references} one word in main memory. LSQ Lookahead, at a high level, looks ahead in the processor's load/store queues\footnote{\tsub{The LSQ stores the necessary metadata (e.g., register destination identifiers, operands, and virtual and physical addresses of memory operands) to correctly execute and commit load and store instructions. We refer the readers to~\cite{smith1987zs1,yeager1996mips,smith1997microarchitecture} for implementation details of the LSQ in modern microprocessors.}} and finds load and store instructions that reference \emph{different words/sectors in the same cache block}. LSQ Lookahead then collects the word/sector references, made by younger load/store instructions to the same cache block as the oldest load/store instruction, and stores the collected sector references in the oldest load/store instruction's sector bits. This way, a load/store instruction, when executed, retrieves all words in a cache block that will be referenced in the near future (by younger load/store instructions) to the L1 cache with \emph{only} one cache access.}

{Fig.~\ref{fig:simple-prediction} depicts how LSQ Lookahead is implemented over an example using load instructions. We extend each load address queue (LAQ, \atbcr{1}{stores metadata for load instructions}) entry with sector bits (SB).}
\new{LSQ Lookahead works in two steps. First, when a new entry is allocated at LAQ's tail (\dingOne{}), the LSU compares the new entry's cache block address (CB address) with each of the existing entries' cache block addresses (\dingTwo{}). Second, when it finds a matching cache block address, it updates the existing entry's sector bits by setting the bit that corresponds to the word referenced by the new entry (\dingThree{}).
}

\noindent
\new{\textbf{Sector Predictor (SP).}}
{Although LSQ Lookahead prevents some of the sector misses, it alone \emph{cannot} significantly reduce the number of sector misses. This is because \atbcr{1}{load/store queues are typically \emph{not} large enough to store many load/store instructions, and dependencies (e.g., data dependencies) prevent the processor core from computing the memory addresses of future load/store instructions}. {Thus, we require a more powerful mechanism to complement LSQ Lookahead and minimize sector misses. \new{To this end, we develop the Sector Predictor (SP).}}}

\revdel{We develop the Sector Predictor (SP) to further alleviate sector misses and reduce the amount of high-latency memory accesses in \X{}.} \omcr{1}{SP, at a high level, records which words are used while a cache block is in the cache. The next time the same cache block misses, SP uses that signature to predict that the load would need the same words.} SP \atbcr{1}{leverages two key observations to accurately predict which words a load needs to access}. \atbcr{1}{First, the processor will ``touch'' (access or update) one or multiple words in a cache block from the moment a cache block is fetched to system caches (from main memory) until it is evicted to main memory. The touched words in a cache block \emph{will likely be touched again} when the cache block is next fetched from main memory. Second, dynamic instances of the same static load/store instruction likely touch the same words in different cache blocks. For example, a static load/store instruction in a loop may perform strided accesses to the same word offset in \omcr{2}{different} cache blocks}. \new{SP builds on a class of predictors referred to as spatial pattern predictors (e.g.,~\cite{kumar1998exploiting,chen2004accurate}). We tailor \new{SP} for predicting a cache block's useful words \omcr{1}{(those that are referenced by the processor during the cache block's residency in system caches)}, similar to what is done by~\cite{yoon2012dynamic}}.

{Fig.~\ref{fig:sector-predictor} depicts the organization of the SP. The Sector History Table (SHT) stores the \emph{previously used sectors} that \atbcr{1}{identify the sectors (words) that were touched by the processor in a now evicted cache block in the L1 cache (\dingOne{}).} 
\new{SHT} is accessed with a \emph{table index} that is computed by XOR-ing parts of the \atbcr{1}{load/store instruction's address} with the \emph{word offset} \atbcr{1}{of the load/store instruction's memory address} upon an L1 cache miss (\dingTwo{}).}
We extend the L1 cache to store the table index and \atbcr{1}{the} \emph{currently used sectors} (\dingThree{}). The {currently used sectors} in the cache track which sectors are used \new{during a cache block's residency\revdel{in the L1 cache}}. The table index is used to update the previously used sectors \new{in an entry in} the SHT with the currently used sectors stored in the cache block \atbcr{1}{upon the cache block's eviction} (\dingFour{}).\revdel{ Next, we briefly describe how the system manages the SP.}

\begin{figure}[h]
    \centering
    \begin{subfigure}{0.45\textwidth}
    \includegraphics[width=1.0\textwidth]{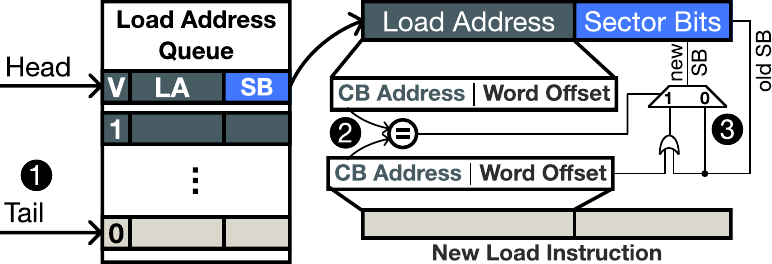}
    \caption{LSQ Lookahead mechanism}
    \label{fig:simple-prediction}    
    \end{subfigure}
    \begin{subfigure}{0.45\textwidth}
    \includegraphics[width=1.0\textwidth]{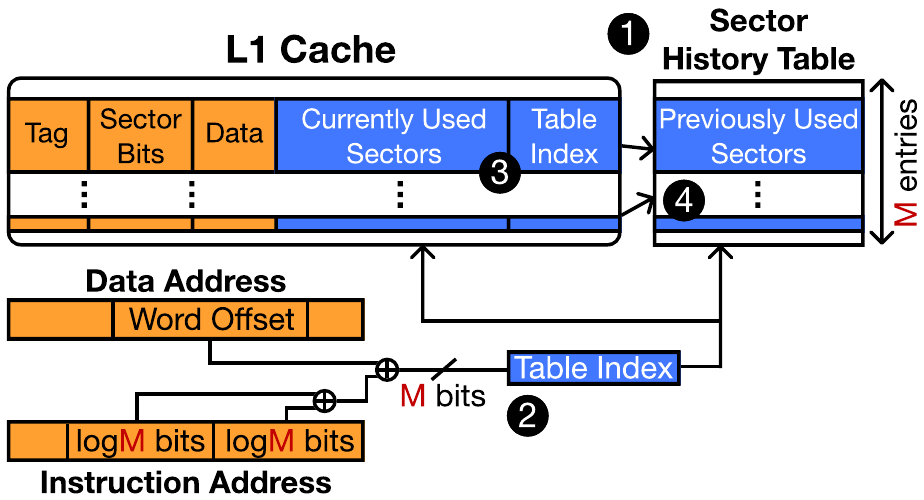}
    \caption{Sector Predictor table and indexing}
    \label{fig:sector-predictor}
    \end{subfigure}
    \caption{\atbcr{2}{Two system components of \X{}: a)} LSQ Lookahead and \atbcr{2}{b)} Sector Predictor}
\end{figure}

\atbcr{1}{We describe how SP operates (not shown in the figure) in five steps based on an example where a memory request accesses the L1 cache. First, when the} {memory request causes a cache miss or a sector miss, the SHT is queried with the table index to retrieve the {\tsub{previously} used sectors}. \atbcr{1}{Second, the previously used \omcr{2}{sectors}} are added to the sector bits of the memory request and forwarded to the next level \nis{in the memory hierarchy}. \tsub{\atbcr{1}{Third,} the L1 miss allocates a new cache block in the L1 cache.} \atbcr{1}{Fourth, t}he table index of the newly allocated cache block is updated with the table index used to access the SHT, and the cache block's currently used sectors are set to logic-0. \atbcr{1}{Fifth, o}nce the \atbcr{1}{missing} cache block is placed in the L1 cache, \new{the cache block's} currently used sectors \tsub{start tracking} the words that are \atbcr{1}{touched} by \atbcr{1}{future} load/store instructions. When the same \atbcr{1}{cache} block is evicted \nis{from the L1 cache}, the SHT entry corresponding to the cache block's table index \tsub{is updated} with the currently used sectors.}\footnote{\atbcr{1}{Due to aliasing in SHT index computation, multiple cache blocks can point to the same SHT entry. SHT receives conflicting updates if two or more such cache blocks are evicted from L1 in the same clock cycle. SHT selects and applies only one update among conflicting updates.}}

\section{Evaluation Methodology}
\label{sec:methodology}
\gf{We describe the workloads  (\cref{sec:methodology:workloads}), power model (\cref{sec:methdology:power}),
and\revdel{ our} simulation infrastructure used to evaluate \X{}\revdel{'s performance and energy} (\cref{sec:methodology:simulation}).} \Copy{R4/3}{\gf{Table~\ref{table:system-configuration} shows our system configuration \tacorevd{that we simulate using Ramulator~\cite{kim2016ramulator,ramulator.github,luo2023ramulator,ramulator2github}}. \tacorevd{Ramulator implements all standard DDR4 timing parameters.}}}\changev{\ref{q:r4q1} and \ref{q:r4q3}} \atbcr{1}{Our simulation infrastructure is open source~\cite{self.github}.}

\subsection{\new{Workloads}}
\label{sec:methodology:workloads} 

{We use 41 workloads from the SPEC2006~\cite{spec2006} (23 workloads), SPEC2017~\cite{spec2017} (12 workloads)\atbcr{1}{\footnote{\atbcr{1}{Due to our trace generation infrastructure limitations, we evaluate \emph{only} those workloads that we could successfully generate traces for.}}}, and DAMOV~\cite{oliveira2021damov} (6 \atbcr{1}{DRAM-bandwidth-bottlenecked representative application functions}) to evaluate \X{}. For every workload, we generate memory traces corresponding to 100 million instructions from representative regions in each workload using \nis{SimPoint}~\cite{hamerly2005simpoint}. \gf{We classify the workloads into three memory intensity categories \atbcr{1}{(as also done by prior work~\cite{oliveira2021damov,hashemi.isca16}), which Table~\ref{table:workloads} describes, using their observed last-level-cache (LLC) misses-per-kilo-instruction (MPKI)}.}

\begin{table}[ht]
\tempcommand{0.9}
\Copy{R4/1table}{
\scriptsize
\caption{System configuration}
\centering
\resizebox{\linewidth}{!}{
\begin{tabular}{ll}
\hline
\multirow{4}{*}{\textbf{Processor}} & 1--\new{16} cores, 3.6 GHz clock frequency, 4-wide issue\\
& 8 MSHRs per core, 128-entry issue window\\
& 32 KiB L1, 256 KiB L2, 8 MiB L3 caches\\
& \gf{Dynamic Power: 101.7 W~\cite{mcpat}, Static Power: 32.0 W~\cite{mcpat}}\\
\hline
\multirow{4}{*}{\textbf{Mem. Ctrl.}} & 64 entry read/write request queue, FR-FCFS-Cap~\cite{mutlu2007stall}\\
& scheduling policy, Open-page row buffer policy,\\
& \tacorevd{Auto-precharge on last read/write to a row}\\
& \microrev{Row-Bank-Rank-Column-Channel address mapping}~\cite{kim2016ramulator}\\
\hline
\multirow{5}{*}{\textbf{DRAM}} & DDR4~\cite{jedec2017ddr4}, \tacorevd{3200 MT/s data transfer rate}, \tacorevd{1, 2, and 4 channels}\\
& 4 ranks, 16 banks/rank, {32K rows/bank}\\
& 64 subarrays/bank, 8 sectors/subarray\\
& $tRCD$/$tRAS$/$tRC$/$tFAW$ {13.75/35.00/48.75/25 ns}\\
& \tacorevd{$tRRD\_L$/$tRRD\_S$ {5.00/2.50 ns}}\\
\hline
\multirow{2}{*}{\textbf{\X{}}} & 128-entry LSQ Lookahead (default)\\
& 512-entry Sector Predictor (default)\\
\hline
\end{tabular}
}
\label{table:system-configuration}
}
\hfill
\caption{Evaluated workloads. -20\gf{0}6/-2017 indicates SPEC\revdel{ workloads}
}
\label{table:workloads}
\tempcommand{1.1}
\footnotesize
\centering
\resizebox{\linewidth}{!}{
\begin{tabular}{|l||l|}
\hline
\multicolumn{1}{|c||}{\textbf{LLC MPKI}}                      & \multicolumn{1}{c|}{\textbf{Workloads}}                                                                                                                                                                                                                                                                             \\ \hline \hline
\begin{tabular}[c]{@{}l@{}}$\ge{} 10$ \\ (High)\end{tabular} & \begin{tabular}[c]{@{}l@{}}ligraPageRank, mcf-2006, libquantum-2006, gobmk-2006, \\ ligraMIS, GemsFDTD-2006, bwaves-2006, lbm-2006, \\ lbm   -2017, hashjoinPR\end{tabular}                                                                                                                                            \\ \hline
\begin{tabular}[c]{@{}l@{}}$1.. 10$\\ (Medium)\end{tabular}  & \begin{tabular}[c]{@{}l@{}}omnetpp-2006, gcc-2017, mcf-2017, cactusADM-2006,\\  zeusmp-2006, xalancbmk-2006, ligraKCore, \\ astar-2006, cactus-2017, parest-2017, ligraComponents\end{tabular}                                                                                                                      \\ \hline
\begin{tabular}[c]{@{}l@{}}$\leq 1$\\ (Low)\end{tabular}     & \begin{tabular}[c]{@{}l@{}}splash2Ocean, tonto-2006, xz-2017, wrf-2006, bzip2-2006, \\ xalancbmk-2017, h264ref-2006, hmmer-2006, namd-2017, \\ blender-2017, sjeng-2006, perlbench-2006, x264-2017, \\ deepsjeng-2017, gromacs-2006, gcc-2006, imagick-2017, \\ leela-2017, povray-2006, calculix-2006\end{tabular} \\ \hline
\end{tabular}
}
\end{table}

\noindent
\new{\textbf{Multi-core workloads.} We create 2-, 4-, 8-, and 16-core workloads by replicating the same single-core workload over multiple cores.} We create 16 eight-core workload mixes for each \gf{memory intensity} category by randomly picking eight single-core workloads from every category.

\subsection{Power Model}
\label{sec:methdology:power}

\noindent
\new{\textbf{DRAM power model.} We use the Rambus Power Model~\cite{vogelsang2010understanding,rambuspowermodel} to model a DDR4 \atbcr{2}{chip} (Table~\ref{table:system-configuration}) that supports \X{}.} 
{We modify the model to (i) \revdel{enable a smaller number of local wordlines (i.e., }activate a smaller number of sectors (\XMO{})\revdel{)} and (ii) reduce the burst size of data transfers for partially activated DRAM rows (\XMT{}).} \new{Our model \gf{considers} the power overheads introduced by the sector transistors and\revdel{ the sector} latches.} \Copy{R3/1a}{\tacorevc{Rambus Power Model computes and reports\changev{\ref{q:r3q1}} the current consumed by a sequence of DRAM commands (e.g., $ACT$, $RD$, $WR$, and $NOP$). We use three command sequences described in one major DRAM manufacturer's power calculation guide~\cite{micronddr4powercalculation} to calculate three \atbcr{1}{important current values:} IDD0 ($ACT$), IDD4R ($READ$), and IDD4W ($WRITE$).}}

\noindent
\Copy{R3/2}{\changev{\ref{q:r3q2} and \ref{q:r4q9}}\new{\textbf{Processor power model.} We use an IPC-based model~\cite{yoon2012dynamic,ahn2009future} to estimate the power consumed by the \tacorevc{entire} processor\revdel{ \gf{core}}. 
\gf{The total power our 8-core processor consumes\revdel{, according to the IPC-based power model,} is equal to:}} $\frac{IPC}{4} \times Dynamic\ Power + Static \ Power$. \tacorevc{\atbcr{1}{We comprehensively account for all \X{} power overheads.} Our model includes the dynamic and static power consumed by the sector \new{predictor} and the additional cache storage (modeled by CACTI~\cite{cacti}). Our system energy results represent the energy consumed by main memory and the entire processor during the execution of an evaluated workload.}}

\subsection{Performance and Energy}
\label{sec:methodology:simulation}

\Copy{R3min/4}{We evaluate \X{}'s performance and \gf{energy} using \nis{a modified version of Ramulator~\cite{kim2016ramulator,luo2023ramulator}, a cycle-accurate, trace-based DRAM simulator, and a modified version of DRAMPower~\cite{chandrasekar2012drampower}, a DRAM power and energy estimation tool}. 
\new{\gf{We extend Ramulator by implementing \X{}'s LSQ Lookahead and Sector Predictor \tsub{as described in \cref{sec:system-integration}}. \finallabel{\ref{q:r3q4}}\tsub{We modify how Ramulator enforces the \tFAW{} timing constraint. Our modification allows for 32 sectors (i.e., the number of sectors in four rows) to be activated within a \tFAW{} (e.g., the memory controller can activate 4 sectors from 8 different rows and 8 sectors from 4 different rows).} \tacofinalc{The rate of $ACT$ commands is still constrained by the $tRRD\_{L}$ and $tRRD\_{S}$ parameters in our modified Ramulator model. Therefore, our memory controller issues up to 10 activate commands (calculated as $\SI{25.0}{\nano\second}/\SI{2.5}{\nano\second}$) in a $tFAW$ window.} \tacofinalc{To verify that the peak power draw imposed by the higher rate of finer-granularity $ACT$ commands (e.g., an $ACT$ command that activates \emph{only} one sector from a row) does \emph{not} increase the power requirements of a DRAM chip, we compare the power draw of 1) four activate commands each of which activates all eight sectors in a row (the rate of activate commands are constrained by $tFAW$) to 2) 10 activate commands each of which activates one sector in a row (the rate of activate commands are constrained by $tRRD\_S$), in a default $tFAW$ long time window (\SI{25}{\nano\second} in our configuration) using Rambus Power Model~\cite{rambuspowermodel}. We find that 10 activate commands each of which activates one sector from a row consume {20.34}\% less power than four activate commands each of which activates all eight sectors in a row (and as such \X{} operates within the region that conventional DRAM is designed to operate).} \Copy{R3/1b}{\changev{\ref{q:r3q1}}\tacorevc{We modify DRAMPower by integrating \X{}'s current values (e.g., IDD0, IDD4R, and IDD4W) that we obtain from the \atbcr{1}{modified} Rambus Power Model.}}}}
}

\noindent
\gf{\textbf{Performance metrics.} We measure single-workload performance using \emph{parallel speedup} (i.e., the baseline single-core execution time divided by the multi-core \X{} execution time), which allow\new{s} us to evaluate \X{}'s scalability for a single workload. We measure workload mix performance using \emph{weighted speedup}~\cite{snavely2000symbiotic}, which allows us to evaluate \X{}'s system throughput~\cite{eyerman2008system} in a heterogeneous computing environment.}

\section{\gf{Evaluation} Results}

\gf{We evaluate \X{}'s impact on DRAM power, LLC MPKI, performance, energy, and DRAM area.}

\subsection{\gf{Impact on DRAM Power}}
\label{sec:results-dram-power}

\gf{Fig.}~\ref{fig:idd-plot} \gf{shows \X{}'s impact on DRAM power consumption. We analyze} the \gf{DRAM array power, DRAM peripher\omcr{1}{al} circuitry} power, \atbcr{3}{and DRAM energy consumed by \X{}} to perform $ACT$, $READ$, and $WRITE$ \gf{DRAM} operations \gf{for 8, 4, 2, and 1 sectors.} \atbcr{1}{Our results show that $READ$ and $WRITE$ power \atbcr{3}{and energy} greatly reduces as fewer sectors are read or written to.}

\begin{figure*}[!th]
    \centering
    \includegraphics[width=1.0\linewidth]{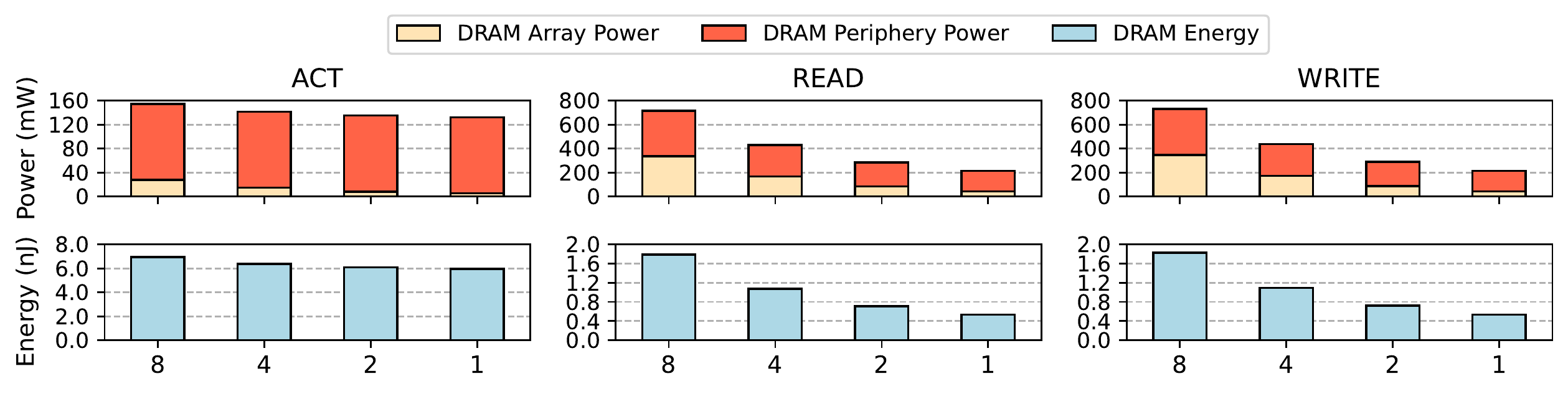}
    \caption{\revdel{$ACT$/$READ$/$WRITE$}\new{DRAM command} power \atbcr{3}{and energy} for varying number of sectors}
    \label{fig:idd-plot}
\end{figure*}

We make three observations \atbcr{1}{from Fig.~\ref{fig:idd-plot}}. 
\atbcr{1}{First,} \XMO{}\revdel{together with} and \XMT{} significantly improve $READ$ and $WRITE$ power consumption. We find that the power consumed by DRAM while reading from and writing to a sector is 70.0\% and 70.6\% smaller than reading from and writing to all sectors, respectively. This improvement is \gf{due to the} 
(i) reduced sense amplifier activity in \gf{the} DRAM \gf{array}, 
(ii) reduced switching on the \gf{DRAM} peripher\omcr{1}{al} circuitry that transfers data between the \gf{DRAM} array and the \gf{DRAM} I/O, and 
(iii) smaller number of \atbcr{1}{beats in a burst} to transfer data between the DRAM module and the memory controller. 
 
\atbcr{1}{Second,} activating \gf{only} one sector greatly reduce\new{s} the power consumed by the DRAM array compared to activating all eight sectors. Because \XMO{} enables activating a small set of DRAM sense amplifiers in a DRAM row\revdel{($\frac{1}{8}$ of all sense amplifiers in the row)}, activating a single sector consumes 66.5\% less \gf{DRAM array} power compared to activating eight sectors. \new{However, w}e find that activating one sector reduces the overall power consumption of an $ACT$ operation by \new{only} 12.7\% \gf{compared to the baseline DDR4 \new{module}}. Th\new{is} effect \revdel{of reducing activation power on overall power consumption}is \new{small} since the power consumed by the peripher\omcr{1}{al} circuitry makes up a large proportion of the activation power and is not affected by the number of sectors activated.
\gf{Third,} the \revdel{additional }circuitry required to implement \XMO{} incurs little activation power overhead. Compared to the baseline DDR4 module, \XMO{} increases activation power by only 0.26\% due to additional switching activity in master wordline drivers \revdel{from the sector transistors used to control the activation of each sector}(\cref{sec:sectored-activation-mechanism}).

\noindent
\atbcr{1}{\textbf{Effects of DRAM Bus Frequency.}} \changev{\ref{q:r4q5}}\Copy{R4/5}{\tacorevd{We investigate how DRAM bus frequency affects the read power (IDD4R) relative to the activate power (IDD0). We repeat our experiments using 2$\times{}$ the frequency of the baseline bus frequency (i.e., 3200 MHz or 6400 MT/s). The read power (IDD4R) is $12.39\times$ and $12.42\times$ higher than the activate power (IDD0) at the baseline bus frequency (3200 MT/s) and twice the baseline bus frequency (6400 MT/s), respectively (this observation is in line with that of a prior work's~\cite{david2011memory}). 

We conclude that \atbcr{1}{1) \X{}'s fine-grained DRAM \new{data transfer and activation} provide a significant reduction in DRAM read and write power \atbcr{3}{and energy} and} 2) the bus frequency does \emph{not} significantly affect DRAM read power relative to DRAM activate power.}}

\revdel{\gf{We conclude that \X{}'s fine-grained DRAM \new{access and activation} provides a significant reduction in DRAM's power consumption. }}

\subsection{\atbcr{1}{Number of} Sector Misses}

{To quantify the number of sector misses\revdel{(recorded when a cache block is present, but some sectors are missing, } (see \cref{sec:tracking-valid-words}), we look at the LLC MPKI of workloads \atbcr{2}{run} with \new{different} \X{} configurations.} Fig.~\ref{fig:sector-misses} plots the LLC MPKI {for different LSQ Lookahead (LA$<$\nis{\emph{number}}$>$ where \nis{\emph{number}} is the number of entries looked ahead in the LSQ) and Sector Predictor (SP$<$\nis{\emph{number}}$>$ where \nis{\emph{number}} is the number of entries in the SHT)} {configurations along with the \emph{Basic} \X{} configuration which does not use LSQ Lookahead \new{nor} SP}. Each bar shows the \atbcr{1}{average LLC MPKI across all evaluated workloads in each benchmark suite (x-axis, see Table~\ref{table:workloads} for a list of all workloads classified according to their LLC MPKI) for a \X{} configuration}. 

\begin{figure*}[!th]
    \centering
    \includegraphics[width=1.0\linewidth]{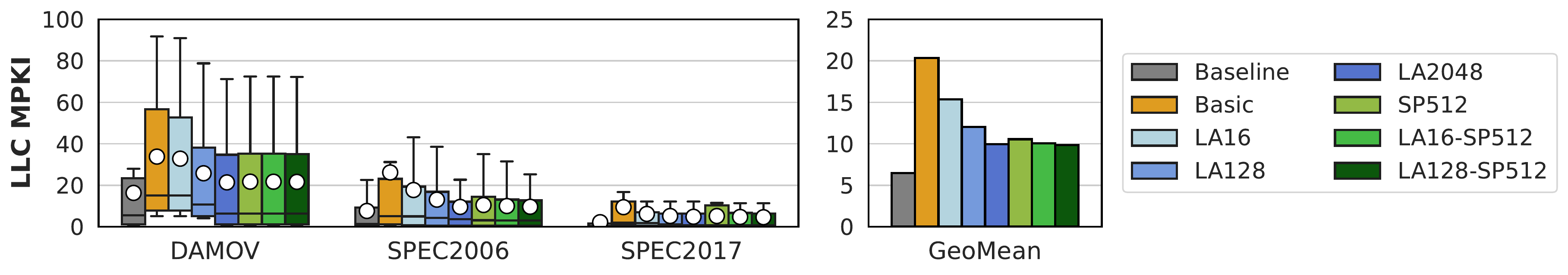}
    \caption{LLC MPKI for different \X{} configurations}
    \label{fig:sector-misses}
\end{figure*}

We make three major observations\revdel{from Fig.~\ref{fig:sector-misses}}. {First, \atbcr{1}{\X{} without LSQ Lookahead and SP (Basic)} greatly increases the LLC MPKI of a workload, on average by \omcr{1}{3.1}$\times{}$\new{, compared to the baseline,} due to sector misses. Second, LSQ Lookahead reduces the number of LLC misses \atbcr{1}{of Basic} by \omcr{1}{25}\%, \omcr{1}{41}\%, and \omcr{1}{51}\% by looking ahead 16, 128, and a currently \omcr{2}{very costly} 2048 younger entries in the LSQ, respectively. This is because LSQ Lookahead can identify the words that will be used in a cache block and retrieve these from DRAM with a single memory request. Third, LSQ Lookahead together with \new{SP} (LA128-SP512) reduces the number of LLC misses \new{of \atbcr{1}{Basic}} by \omcr{1}{52}\% \atbcr{1}{and of LA128 by \omcr{1}{18}\%.} \omcr{1}{LA128-SP512} performs as well as the currently  \omcr{2}{very costly} implementation of LSQ Lookahead \atbcr{1}{which looks ahead} 2048 entries \atbcr{1}{in the LSQ}. \atbcr{1}{LA128-SP512 does so as SP greatly reduces the number of additional LLC misses by recognizing} intra-cache-block access patterns from previously \atbcr{2}{performed} memory requests and correctly predicts the words that will be used \atbcr{1}{in a cache block}.}

\new{We conclude that LSQ Lookahead with a 128 lookahead size together with SP minimizes the LLC misses caused by sector misses. We use the LA128-SP512 configuration in the remainder of our evaluation.}

\subsection{\new{Single-Workload Performance and Energy}}
\label{sec:perf-single}



\new{We evaluate \X{}'s performance and energy using (i) single-core workloads, and (ii) 2-, 4-, 8-, and 16-core multi-programmed workloads\revdel{that are} made up of identical single-core workloads. We compare \X{}'s performance and system energy to a baseline coarse-grained DRAM system.}

\noindent
\new\textbf{\tacorevcommon{Microbenchmark performance.}}
\tacorevcommon{Fig.~\ref{fig:microbenchmark-performance} shows the normalized parallel speedup of a random access (Random, left) and a strided streaming access (Stride, right) workload for the baseline system and \X{} \atbcr{1}{(LA128-SP512)}. The Random workload accesses i) one randomly determined word in main memory (8 bytes) by executing a load instruction every five instructions and ii) has a very high LLC MPKI of 178.29. These two properties of Random make it a good fit for \X{} as Random accesses \emph{only} one sector in every cache line. The Stride workload accesses i) every word address in a contiguous, 16 MiB large memory address range with a stride of 64 bytes, i.e., Stride accesses the following addresses [0, 64, 128, ..., 8, 72, 132, ..., 16, ...] and ii) has a very high LLC MPKI of 78.57. Stride is a \omcr{1}{poor} fit for \X{} because every access to a word in a cache line results in a \emph{sector miss} (\atbcr{1}{none of the} accessed 8-byte \atbcr{1}{words are} cached and \atbcr{1}{these words \atbcr{2}{have}} to be fetched from main memory)\atbcr{1}{, where the large cache block reuse distance prevents LSQ Lookahead from prefetching all useful words in a cache block}.}
\begin{figure*}[!h]
    \centering
    \includegraphics[width=\textwidth]{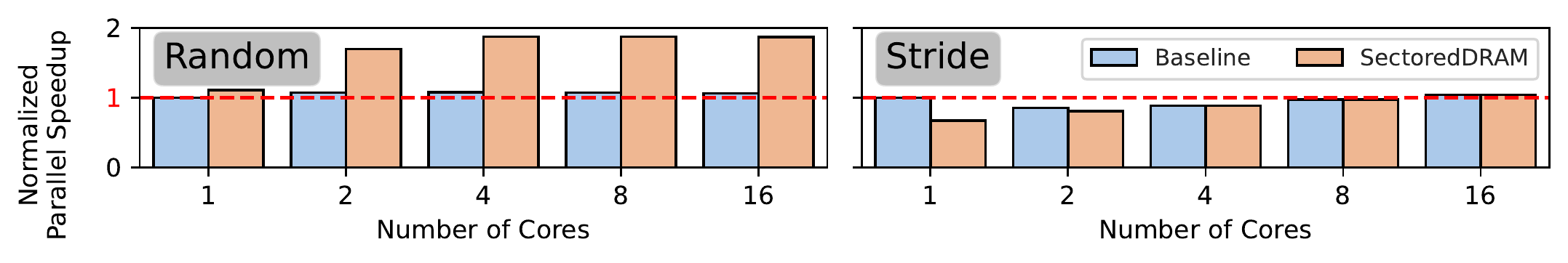}
    \caption{\tacorevcommon{Normalized parallel speedup of random and strided streaming microbenchmarks for varying number of cores}}
    \label{fig:microbenchmark-performance}
\end{figure*}

\tacorevcommon{We make two major observations from Fig.~\ref{fig:microbenchmark-performance}. First, \X{} provides significant performance benefits for workloads that randomly access words (e.g., Random). \X{}'s performance benefits increase with the number of cores (i.e., with increasing LLC MPKI) for Random because a larger fraction of all memory requests (random word accesses) benefit from \X{}'s reduction in $tFAW$ (\cref{sec:sectored-activation-mechanism}). \X{} provides 1.11$\times{}$, 1.69$\times{}$, 1.87$\times{}$, 1.87$\times{}$, and 1.87$\times{}$ normalized parallel speedup for 1, 2, 4, 8, and 16 cores, respectively for Random. Second, \X{} reduces system performance for workloads that frequently cause sector misses (e.g., Stride). \X{} provides 0.67$\times{}$, 0.95$\times{}$, 1.00$\times{}$, 1.00$\times{}$, and 1.00$\times{}$ the normalized parallel speedup of Baseline for 1, 2, 4, 8, and 16 cores, respectively for Stride. \X{}'s performance becomes closer to Baseline as the number of cores increases. This is because the LLC is \emph{not} large enough to store all cache lines accessed by four or more cores for Stride (i.e., Baseline accesses main memory to retrieve each word, similarly to \X{}).}

\noindent
\textbf{Performance.} 
The two lines in Fig.~\ref{fig:parallel-speedup} show the normalized parallel speedup (on the primary/left y-axis) of three \gf{representative} high MPKI (top row), medium MPKI (middle row), and low MPKI (bottom row) workloads for the baseline system (solid lines) and \X{} (dashed lines). Fig.~\ref{fig:box_plot} (top row) shows the distribution of normalized parallel speedups of all high, medium, and low MPKI workloads.\footnote{\label{footnote:box}\gf{Each box is lower-bounded by the first quartile and upper-bounded by the third quartile. The median falls within the box. The inter-quartile range (IQR) is the distance between the first and third quartiles (i.e., box size). Whiskers extend to the minimum and maximum data point values on either side of the box, while a bubble depicts average values.}} 

\begin{figure*}[!ht]
    \begin{subfigure}[!h]{0.46\textwidth}
    \includegraphics[width=\textwidth]{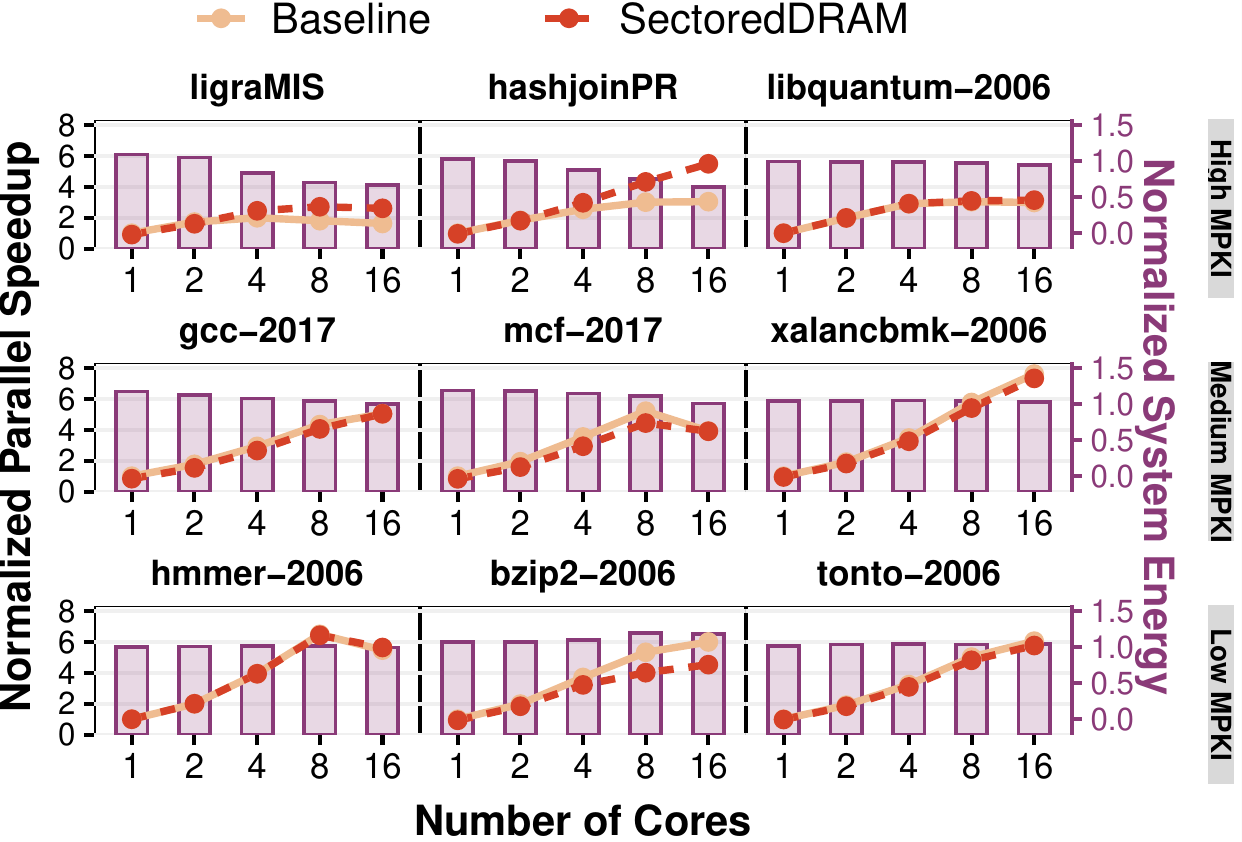}
    \caption{\gf{Normalized parallel speedup (primary/left y-axis) and system energy (secondary/right x-axis) of representative high, medium, and low LLC MPKI workloads for 1--16 cores}}
    \label{fig:parallel-speedup}
    \end{subfigure}
    \hfill
    \begin{subfigure}[!h]{0.46\textwidth}
    \includegraphics[width=\textwidth]{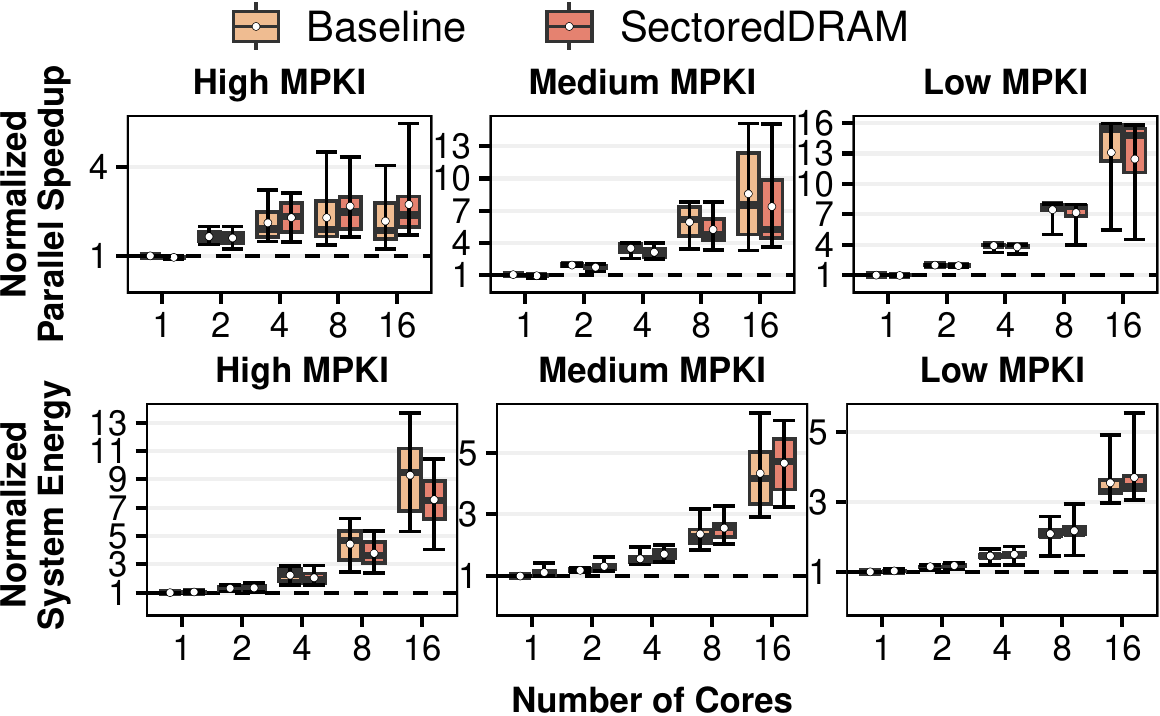}
    \caption{\tacoreva{Normalized parallel speedup (top) and system energy (bottom) distribution of all high, medium, and low LLC MPKI workloads for varying number of cores}\footref{footnote:box}}
    \label{fig:box_plot}        
    \end{subfigure}
    \caption{\X{} system performance and energy}
\end{figure*}




{We make \tacorevd{three} observations from the two figures. First, \X{} provides higher parallel speedup over the baseline for high MPKI workloads when the number of cores is larger than two. For example, \X{} provides \gf{26\%} higher parallel speedup than the baseline for \gf{all} 16-core high MPKI workloads on average.\changev{\ref{q:r1q1}}\footnote{\Copy{R1/1}{\tacoreva{We observe a decrease in normalized parallel speedup for hmmer-2006 and mcf-2017 as the number of cores increase from 8 to 16. This is because the two workloads, \atbcr{2}{when ran} on 16 cores, contend for memory at a higher degree than \atbcr{1}{when ran} on 8 cores in the simulated system. For example, 16-core hmmer-2006 has 24.41\% higher average latency for memory requests than 8-core hmmer2006.}}} As the average row buffer hit rate for 16-core high MPKI workloads is only 18\%, the memory controller needs to issue many $ACT$ (activate) commands to serve the memory requests. \X{}'s \tFAW{} reduction \gf{(\cref{sec:sectored-activation-mechanism})} allows the memory controller to issue the large number of $ACT$ commands required by these workloads \gf{(i.e., 82\% of all main memory requests)} at a higher rate, reducing the average memory \gf{access} latency for these workloads \gf{(by 25\% on average for 16-core  high MPKI workloads)}.
\revdel{The parallel speedup of two displayed applications, \emph{ligraPageRank} and \emph{hashjoinPR}, follow the common pattern we describe for the high MPKI workloads. However, \emph{libquantum} behaves differently where \X{} cannot provide substantially higher parallel speedups than the baseline as the number of cores increase. This workload has a high row buffer hit rate (62\% for the 16-core \emph{libquantum}). Thus, the memory controller does \emph{not} need to issue many $ACT$ commands to serve \emph{libquntum}'s memory requests, leaving a small window of opportunity for the memory controller to schedule $ACT$ commands at a higher rate provided by the \tFAW{} reduction.}Second, \X{}, on average, provides a smaller parallel speedup compared to the baseline for low and medium MPKI workloads. Although \X{}'s \tFAW{} reduction reduces the proportion of processor cycles where the memory controller has to stall to satisfy the \tFAW{} timing parameter from 14.4\% in the baseline to 6.5\% in \X{} for 16-core low and medium MPKI workloads, the average memory latency for these workloads increases by \gf{0}.5\% in \X{} compared to the baseline. Moreover, for these workloads, sector misses increase the number of memory requests on average by 69\%. Because a larger number of memory requests experience higher latencies in \X{} compared to the baseline, \X{} provides a smaller parallel speedup for these workloads.} 
\changev{\ref{q:r4q8}}\Copy{R4/8}{\tacorevd{Third, \X{} incurs 5.41\% performance overhead on average across all single-core workloads. We attribute this to sector misses that increase the number of memory requests and the average memory latency.}}

\noindent
\new{\textbf{System energy consumption.} The bars in Fig.~\ref{fig:parallel-speedup} show the system energy consumption (on the secondary/right y-axis) of \X{} normalized to the system energy consumption of the baseline. Fig.~\ref{fig:box_plot} \gf{(bottom)} shows the distribution of normalized system energy consumption for workloads from three categories (low, medium, and high MPKI). We make two observations. \changev{\ref{q:r1q3}}\Copy{R1/3}{\tacoreva{First, \X{} reduces system energy consumption for high MPKI workloads when the number of cores is larger than two. On average at 16 cores, high MPKI workloads' system energy consumption reduces by \gf{20\%}. \X{} achieves this by a combination of (i) reduced DRAM energy consumption due to \XMO{} and \XMT{} {(we present a detailed breakdown of \XMO{} and \XMT{}'s effects on DRAM energy consumption in}~\cref{sec:results-perf-energy}), and (ii) reduced background power consumption by the \atbcr{1}{computing system} as workloads execute faster. Second, for medium and low MPKI workloads, \X{} increases system energy consumption. We observe that \X{} increases the average DRAM energy consumption by 12\% for 16-core medium/low MPKI workloads. The increase in DRAM energy consumption together with the increase in background power consumption by the {1}{computing system} (as workloads execute slower, Fig.~\ref{fig:parallel-speedup}) increases the system energy consumption for these workloads.}}}

\new{We conclude that \X{} improves system performance and reduces system energy consumption in high MPKI workloads \gf{where:
(i) a high number of $ACTs$ targets different DRAM banks (i.e., the workload is bound by \tFAW), and 
(ii) the sector predictor can accurately  predict the used words.}
}

\subsection{\new{\omcr{2}{Multiprogrammed Workload}} Performance and Energy}
\label{sec:results-perf-energy}
\gf{We evaluate \X{}'s performance and energy using multi-programmed workload mixes. To stress DRAM and cache hierarchy, 
{we use {\gf{high} MPKI workload mixes}.}
We compare \X{}'s performance and main memory \gf{access} energy to a baseline coarse-grained DRAM system and \gf{three} state-of-the-art fine-grain\gf{ed DRAM} mechanisms: 
\gf{(i)} Fine-Grained Activation (FGA)~\cite{cooper2010fine,udipi2010rethinking}, 
\gf{(ii)} Partial Row Activation (PRA)~\cite{lee2017partial}, and 
\gf{(iii)} HalfDRAM~\cite{zhang2014half} \atbcr{1}{and HalfPageDRAM~\cite{ha2016improving}}\footnote{\atbcr{1}{We estimate HalfPageDRAM's~\cite{ha2016improving} performance and energy benefits using HalfDRAM~\cite{zhang2014half} as both techniques enable half-size DRAM row activations at high DRAM throughput.}}. \atbcr{1}{Unless otherwise stated, HalfDRAM depicts the evaluated performance and energy of both HalfDRAM and HalfPageDRAM in the remainder of the paper.}    }

\noindent
{\textbf{Performance.}} {\gf{Fig.~\ref{fig:multicore-speedup-high-mpki} (top) \gf{shows} the weighted speedup~\cite{das2009application,eyerman2008system,atlas} of \gf{16} workload mixes for \X{} and the three state-of-the-art fine-grained DRAM mechanisms, normalized to the baseline system.} {We make four observations. } \gf{First, \X{}'s weighted speedup is 1.17$\times$ {(1.36$\times$) that of the baseline, on average (maximum)}, across all workload mixes. This is due to \X{}'s \tFAW{} reduction.\revdel{, which allows \X{}'s memory controller to schedule $ACT$ commands at a higher rate than in the baseline. In this way,} \X{} serves $READ$ requests faster (\X{}'s average DRAM read latency is {approximately} 25\% \new{smaller} compared to the baseline) and thus improves the performance of the memory-intensive workloads.} \gf{Second, \X{} greatly outperforms naive fine-grained DRAM mechanisms (i.e., FGA~\cite{cooper2010fine,udipi2010rethinking}). We observe that \X{}'s weighted speedup is 2.05$\times$ that of FGA, on average across all workloads. \new{FGA mechanisms} greatly reduce the throughput of DRAM data transfers as they are limited to fetching a cache block from a single mat (\cref{sec:motivation:limitations}). \revdel{Compared to the baseline, F\new{G}A incurs a 43\% \emph{reduction} in weighted speedup, on average.}} 
\gf{Third, \X{}'s weighted speedup is 1.10$\times$ that of PRA, on average across all workloads. \X{} outperforms PRA by enabling fine-grained DRAM access and activation for \emph{both} $READ$ and $WRITE$ operations, while PRA is limited to $WRITE$ operations.}

\begin{figure}[!h]
    \centering
    \includegraphics[width=1.0\linewidth]{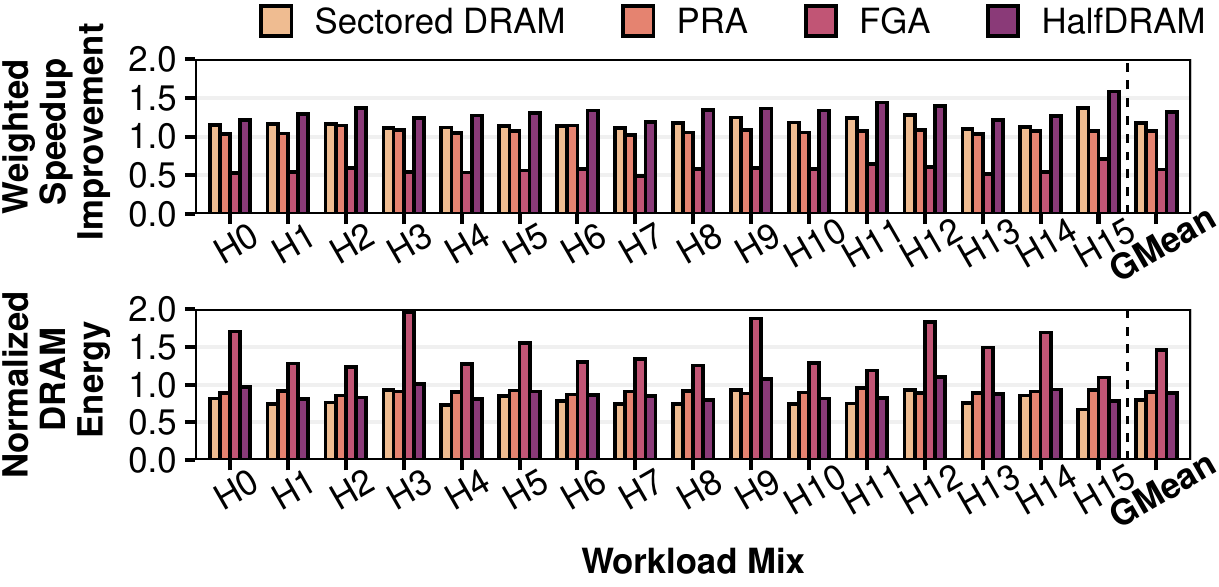}
    \caption{Weighted speedup improvement over the baseline (higher is better) on top. DRAM energy normalized to the baseline (lower is better) on the bottom.}
    \label{fig:multicore-speedup-high-mpki}
\end{figure}

\gf{Fourth, \X{}'s weighted speedup is 0.89$\times$ that of HalfDRAM, on average across all workloads. \X{} \emph{cannot} improve performance as much as HalfDRAM since, in \X{}, the memory controller needs to service additional memory requests caused by sector misses. However, as we show next, HalfDRAM's higher performance benefits come at the cost of \emph{higher} area overheads (\cref{sec:results-area}) and \emph{lower} energy savings than \X{}. 
}}

\noindent
\nis{\textbf{DRAM energy consumption.}} 
\new{\gf{Fig.}~\ref{fig:multicore-speedup-high-mpki} (bottom) \gf{shows} the \gf{DRAM} energy \gf{consumption of each} workload mix for \X{} and \gf{the} state-of-the-art mechanisms. \gf{V}alues are normalized to the DRAM energy in the baseline system.} \new{We observe that (i)} \gf{\X{} \emph{significantly} reduces DRAM energy consumption compared to the baseline, leading to up to (average) {33}\% ({20}\%) lower DRAM energy consumption,} 
\new{and (ii)} \gf{\X{} enables larger DRAM energy savings compared to prior works. On average, across all workload mixes, \X{} reduces DRAM energy consumption by 84\%, 13\%, and 12\% compared to FPA, PRA, and HalfDRAM. \revdel{Note that, while HalfDRAM can provide better performance than \X{}, our mechanism provides higher energy savings than HalfDRAM since it enables \new{accessing} DRAM at \new{a finer granularity}.}}

\gf{We analyze the impact of \X{} on the energy consumed by DRAM operations\revdel{ to better understand our DRAM access energy results}.} {Fig.~\ref{fig:multicore-energy-breakdown} \gf{(left)} \gf{shows} the DRAM energy broken down into $ACT$, background, and $RD/WR$ consumption, normalized to the baseline system DRAM energy consumption, averaged across all workload mixes. We make two observations.} 

\begin{figure}[ht]
    \centering
    \includegraphics[width=\linewidth]{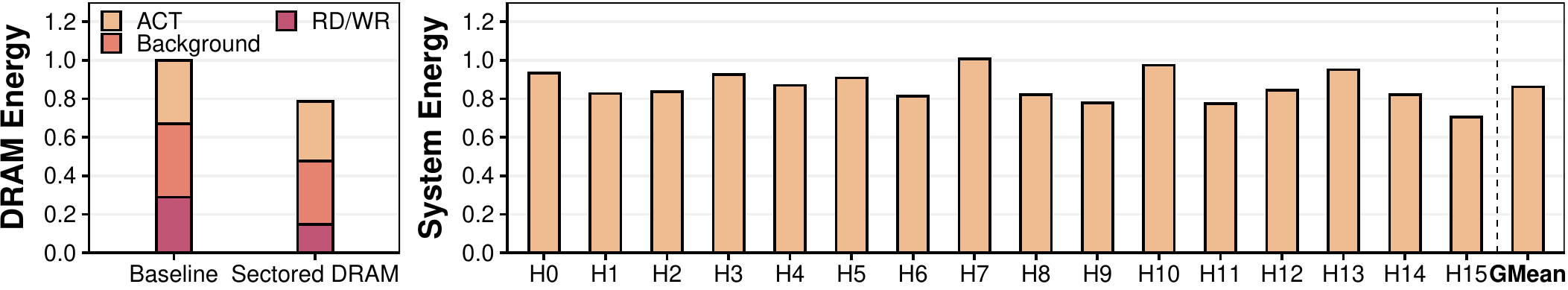}
    \caption{\revdel{Components that contribute to }DRAM energy \new{breakdown} normalized to total DRAM energy consumed by the baseline (left). System energy normalized to the baseline (right) for different workload mixes.}
    \label{fig:multicore-energy-breakdown}
\end{figure}

\gf{First}, \XMechTwo{} (\XMT{}) greatly reduces the $RD/WR$ energy by {51}\%, on average. Using \XMT{}, the system retrieves only the required (and predicted to be required) words of a cache block from the DRAM module. On average, the number of bytes transferred between the memory controller and the DRAM module is reduced by {55}\% (not shown) with \X{} compared to the baseline. In this way, the system uses the power-hungry memory channel more energy-efficiently, eliminating unnecessary data movement. \gf{Second}, \XMechOne{} (\XMO{}) can reduce the energy spent on activating DRAM rows by {6}\% on average. The reduction in $ACT$ energy is relatively small. This is because the memory controller issues \new{more} $ACT$ commands compared to the baseline in \X{}. The new $ACT$ commands (i) respond to the \new{additional} memory requests caused by sector misses, and (ii) resolve the row conflicts that occur due to interference created by the sector misses.

\noindent
\nis{\textbf{System energy consumption.}} {\gf{Fig.}~\ref{fig:multicore-energy-breakdown} \gf{(right) shows} the energy consumed by the \X{} system (processor \emph{\gf{and}} DRAM) normalized to the energy consumed by the baseline system for all workloads. We observe that \X{} reduces system energy consumption, on average {(at maximum), by {14}\% (23\%)}. \X{} does so by (i) reducing DRAM energy consumption, and (ii) reducing background power consumption by the processor as workloads execute faster.} 

\revdel{\gf{We conclude that \X{} improves performance and provides energy savings compared to current coarse-grained DRAM systems and prior state-of-the-art fine-grained DRAM systems.}}

\subsection{Area Overhead} 
\label{sec:results-area}
\noindent
\textbf{Modeled DRAM chip.} \new{We use CACTI~\cite{cacti} to model the area of a DRAM chip (Table~\ref{table:system-configuration}) using $22nm$ technology. \atbcr{1}{Our model is open source~\cite{self.github}}.} 
\gf{{Our modeled DRAM chip takes up, in each bank}:
(i)~\SI{8.3}{\milli\meter\squared} for DRAM cells,
(ii)~\SI{3.2}{\milli\meter\squared} for wordline drivers,
(iii)~\SI{4.6}{\milli\meter\squared} for sense amplifiers,
(iv)~\SI{0.1}{\milli\meter\squared} for row decoder,
(v)~$<$\SI{0.1}{\milli\meter\squared} for column decoder, and
(vi)~$<$\SI{0.4}{\milli\meter\squared} for data and address bus.
}

\revdel{
\begin{table}[h] 
    \caption{DRAM bank area breakdown.}
    \centering
    \footnotesize
    \begin{tabular}{  l r }
    \textbf{DRAM Component} & \textbf{Area ($mm^2$)} \\
    \hline
    DRAM cells & $8.3$ \\
    Wordline drivers & $3.2$ \\
    Sense amplifiers & $4.6$ \\
    Row decoder & $0.1$ \\
    Column decoder & $<0.1$ \\
    Data \& address bus & $0.4$ \\ 
    
    \hline
    \end{tabular}
    \label{table:dram-area}
\end{table}
}
\noindent
{\textbf{Sectored DRAM.} \new{We model the overhead of} (i) 8 additional LWD stripes, (ii) sector transistors, (iii) sector latches, and (iv) wires that propagate sector bits from sector latches to sector transistors to implement \XMechOne{} (\cref{sec:sectored-activation-mechanism}).
\X{} introduces $2.26\%$ area overhead {(0.39 $mm^2$)} over the baseline DRAM bank.} {Overall, \X{} increases the area of the chip (16 banks and I/O circuitry) by \emph{only} $1.72\%$.}

\noindent
\new{\textbf{FGA~\cite{cooper2010fine, udipi2010rethinking} and PRA~\cite{lee2017partial}.} We estimate the area overhead of these architectures to be the same as \X{} because they require the same set of modifications to the DRAM array to enable Fine-DRAM-Act.}

\noindent
\new{\textbf{HalfDRAM~\cite{zhang2014half} and HalfPage~\cite{ha2016improving}.} We estimate the chip area overheads of HalfDRAM and HalfPage as {2.6}\% and {5.2}\%, respectively. Both HalfDRAM and HalfPage require 8 additional LWD stripes like \X{} does. HalfDRAM further requires implementing double the number of CSL signals~\cite{zhang2014half} to enable mirrored connection, and HalfPage requires doubling the number of HFFs per mat~\cite{ha2016improving}.}

\noindent
\textbf{Processor.} \finallabel{\ref{q:r3q3}}\Copy{R3min/3}{\new{We use CACTI to model the storage overhead of sector bits in caches {(\tacorevcommon{1} byte\tacorevcommon{/}cache block)} and the sector predictor (1088 bytes\tacorevcommon{/}core). The sector bits \tacofinalc{(200 KiB additional storage for a system with 12.5 MiB cumulative L1, L2, and L3 cache capacity)} and the predictor storage increase the area of the 8-core processor by $1.22\%$.}}

\revdel{\new{We conclude that \X{} can be implemented at low area overhead.}}

\revdel{\subsection{\new{Cache Access Latency}}
\new{Sectored caches~\cite{liptay1968structural, smith1987line, alpert1988performance, hill1984experimental} (\cref{sec:tracking-valid-words}) require minor modifications to existing cache design to enable word-granularity access. Our CACTI simulations show that these modifications increase the L1 cache's access latency from \SI{0.78}{\nano\second} to \SI{0.79}{\nano\second}. This small increase in latency is unlikely to increase the number of processor cycles it takes to access the L1 cache. However, finer-granularity sectored cache support may require more sector bits and increase the additional L1 cache access latency by a greater amount, thereby causing the number of processor cycles it takes to access the L1 cache to increase. To understand the overheads of the increase in processor cycles, we simulate a \X{} design (called SlowCache) with L1, L2, and L3 cache access latency values that are one processor cycle higher than the default \X{} design. SlowCache performs as well as the default \X{} design, providing 17.0\% weighted speedup improvement (compared to the baseline) for the high MPKI workload mixes on average, whereas the default \X{} design provides 17.2\% weighted speedup improvement. We conclude that \X{}'s impact on cache access latency has a negligible effect on \X{}'s system performance improvement.}}

\section{Discussion}
\label{sec:discussion}
\subsection{\tacorevd{\X{} with More Memory Channels}}
\label{sec:channel}

\changev{\ref{q:r4q1}}\Copy{R4/1a}{\tacorevd{We evaluate 2- and 4-channel systems to study \X{}'s impact on memory performance in systems equipped with more than one memory channel. Fig.~\ref{fig:channels} shows the normalized parallel speedup of all high, medium, and low LLC MPKI homogeneous workloads for varying numbers of cores on the x-axis. Different boxes show Baseline and \X{} configurations with 1, 2, and 4 memory channels.}}

\begin{figure*}[!h]
    \centering
    \Copy{R4/1fig}{
    \includegraphics[width=\textwidth]{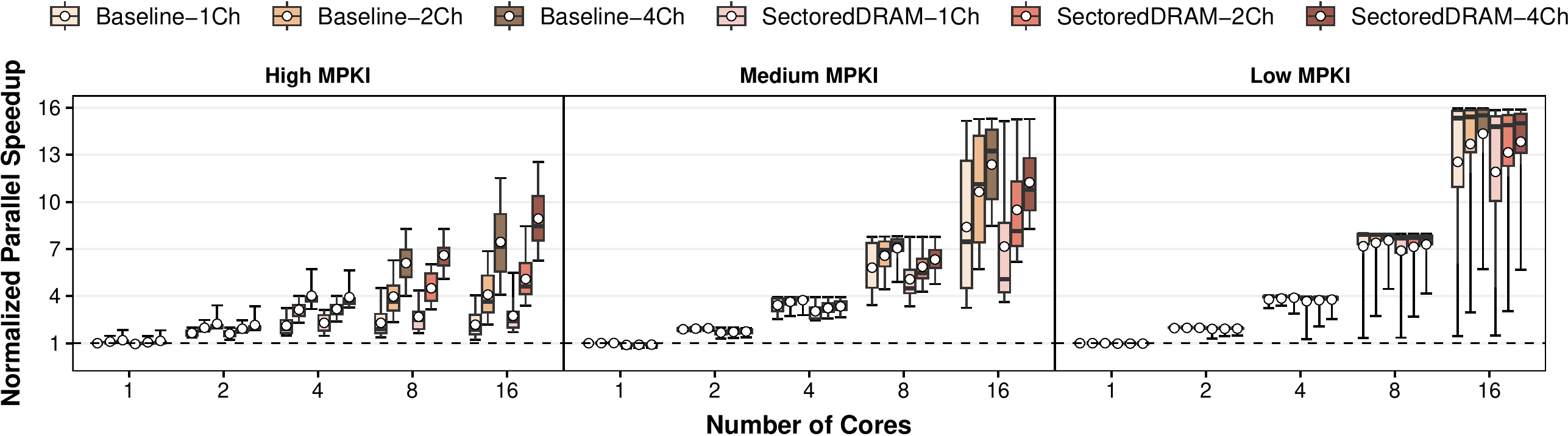}
    \caption{\tacorevd{Normalized parallel speedup of all low, medium, and high LLC MPKI homogeneous workloads for varying number of cores}\footref{footnote:box}}
    \label{fig:channels}
    }
\end{figure*}

\Copy{R4/1b}{\tacorevd{We observe that more memory channels increase system performance on average across all workloads for Baseline and \X{}. For example, for HIGH MPKI 16-core workloads \X{} provides 2.74$\times{}$, 5.09$\times{}$, and 8.94$\times{}$ average normalized parallel speedup for 1-channel, 2-channel, and 4-channel \atbcr{1}{systems}, respectively.}}

\subsection{\nis{Non-Memory-Intensive} Workloads}
\gf{\X{}'s current system integration can lead to varying performance benefits depending on the workload's memory intensity\revdel{, for two reasons}.}
\revdel{{{First,} \gf{\nis{non-memory-intensive} workloads (i.e., workloads with low and medium LLC MPKI; see \cref{sec:methodology:workloads}) do \emph{not} issue enough concurrent memory requests to main memory so \new{they} can benefit from \X{}'s $tFAW$ reduction.}} 
{Second,} the additional memory accesses caused by sector misses \gf{can} increase the average memory \gf{access} latency, causing performance degradation.}
{We propose two approaches to overcome the performance degradation in \nis{non-memory-intensive} workloads.}

\noindent
{\textbf{Dynamically turning \X{} off.} \X{} can be turned off while the system \gf{executes} \nis{non-memory-intensive} \atbcr{1}{(i.e., low and medium MPKI workloads)} workloads. To do so, \gf{we leverage the} performance counters already present in modern processors~\cite{perfmon-events,AMDPerformanceCounter,IntelPerformanceCounter} to periodically compute the average occupancy of the memory controller's read request queue 
{(i.e., the average number of requests in the read request queue)}
and \gf{turn} \X{} \gf{on/off} when the computed value exceeds an empirically determined threshold\revdel{\gf{(}set by the system designer\gf{)}}. {We \gf{(i)} periodically \gf{(every {1000} cycles)} compute the average occupancy \gf{of the memory controller's read request queue}, and 
\gf{(ii)} turn \X{} on for the next {1000} cycles if {the average occupancy} exceeds $30$ \gf{or} turn \X{} off otherwise.}} {\gf{Fig.}~\ref{fig:off-on-dynamic} \gf{shows} the weighted speedup for 16 workload \gf{mixes} from each \gf{memory intensity} category\revdel{(H: high, M: medium, L: low \gf{LLC} MPKI)}\gf{, normalized to the coarse-grained DRAM baseline}. We \gf{show} results for two \revdel{\X{} }configurations: \emph{Always ON} never turns \X{} off and \emph{Dynamic} turns \X{} on and off as described above.}

\begin{figure}[!ht]
    \centering
    \includegraphics[width=1.0\linewidth]{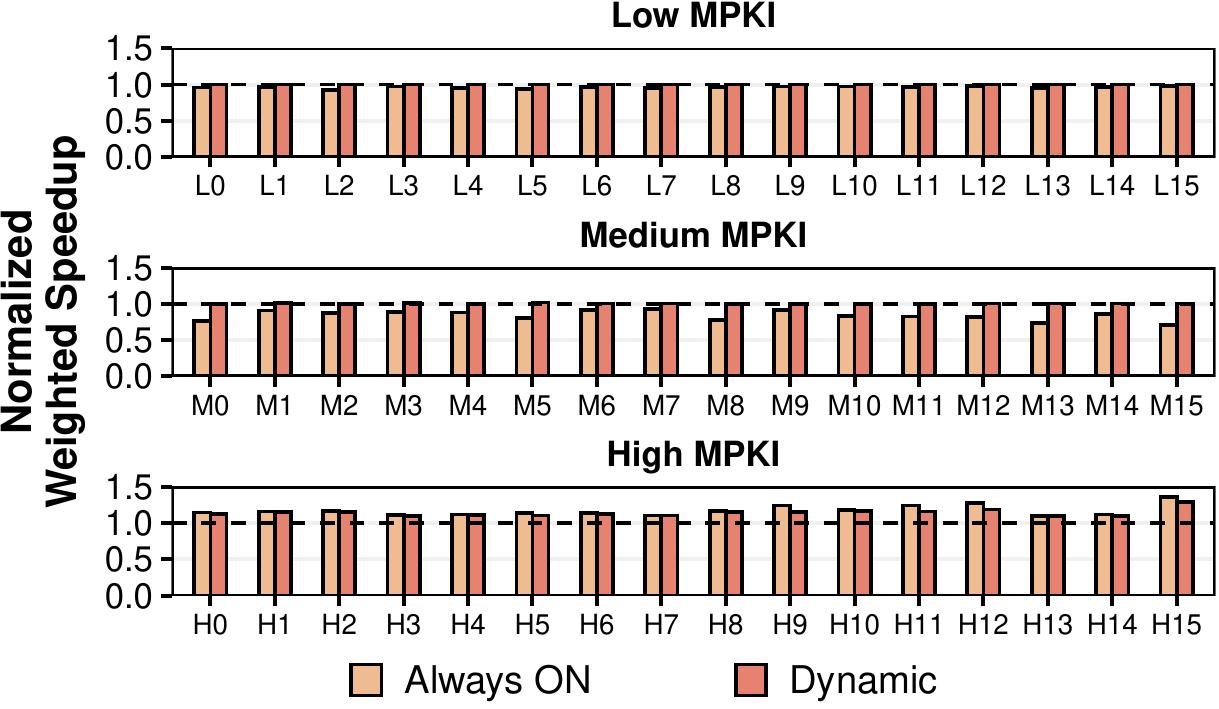}
    \caption{\gf{W}eighted speedup for \emph{Always ON} and \emph{Dynamic}}
    \label{fig:off-on-dynamic}
\end{figure}

\gf{We observe that the \emph{Dynamic} configuration allows \X{} to perform as well as the baseline for non-memory-intensity workloads \new{and} maintain better performance than the baseline for memory-intensive workloads. \emph{Dynamic} provides higher speedups than \emph{Always ON}, on average across all workload mixes and classes, even though \emph{Dynamic} provides slightly lower speedups than \emph{Always ON} for high memory-intensive workloads.}

\noindent
{\textbf{Better sector prediction.} Improving the \new{SP's} accuracy would reduce the\revdel{ number of} additional LLC misses. A more sophisticated SP could be developed by tracking the \revdel{intra-cache-block }access patterns of instructions with deeper history, or other techniques (e.g., reinforcement learning~\cite{bera2021pythia,ipek2008self}, perceptron-based prediction~\cite{jimenez2001dynamic,jimenez2002neural,jimenez2003fast,teran2016perceptron,jimenez2017multiperspective,garza2019bit,bera2022hermes}) could be \new{used} to predict the useful words.} 

\revdel{\new{Implementing SP in the lower level caches (e.g., L2) might improve sector prediction accuracy, potentially at the cost of designing a more complex SP. This is because the access patterns of requests that arrive to the lower level caches could be different than those that arrive to the L1 cache. We implement the SP only in the L1 cache because the L1 cache is typically tightly integrated with the core. This allows us to transfer the memory instruction addresses from the core to the L1 cache at low cost.}}

\subsection{\tacorevcommon{\X{} with Prefetching}}
\label{sec:prefetch}

\tacorevcommon{We implement \X{} support in a simple region-based single-stride prefetcher (based on~\cite{iacobovici2004effective,srinath2007feedback}) to demonstrate \X{}'s performance in a system with prefetching enabled. We model two new system configurations \emph{Baseline-Prefetch} and \emph{\X{}-Prefetch}. Baseline-Prefetch incorporates the simple prefetcher that partitions physical memory address space into 4-KiB (i.e., page-sized) regions and assigns a \emph{stream} to each region. A region is \emph{trained} after four consecutive memory accesses with the same stride, and the prefetcher starts issuing prefetch requests for subsequent last-level cache (LLC) accesses (hits and misses) that target the memory region. We configure the prefetcher to have a degree of four, and the first prefetch request targets four cache blocks ahead of the memory request that accesses the LLC.}
\tacorevcommon{\X{}-Prefetcher augments this prefetcher with support for sector bits. \X{}-Prefetcher sets the sector bits of a prefetch request based on the sector bits of the memory request that resulted in the prefetch request (i.e., hit or missed in the LLC). For example, if a memory request asks for the first three sectors of a cache line, the prefetch request also asks for the first three sectors of their respective cache lines.}

\tacorevcommon{Fig.~\ref{fig:prefetching} shows the normalized speedup (x-axis) for Baseline and \X{} (with and without prefetching) across all evaluated single-core workloads.}

\begin{figure*}[!h]
    \centering
    \includegraphics[width=0.6\linewidth]{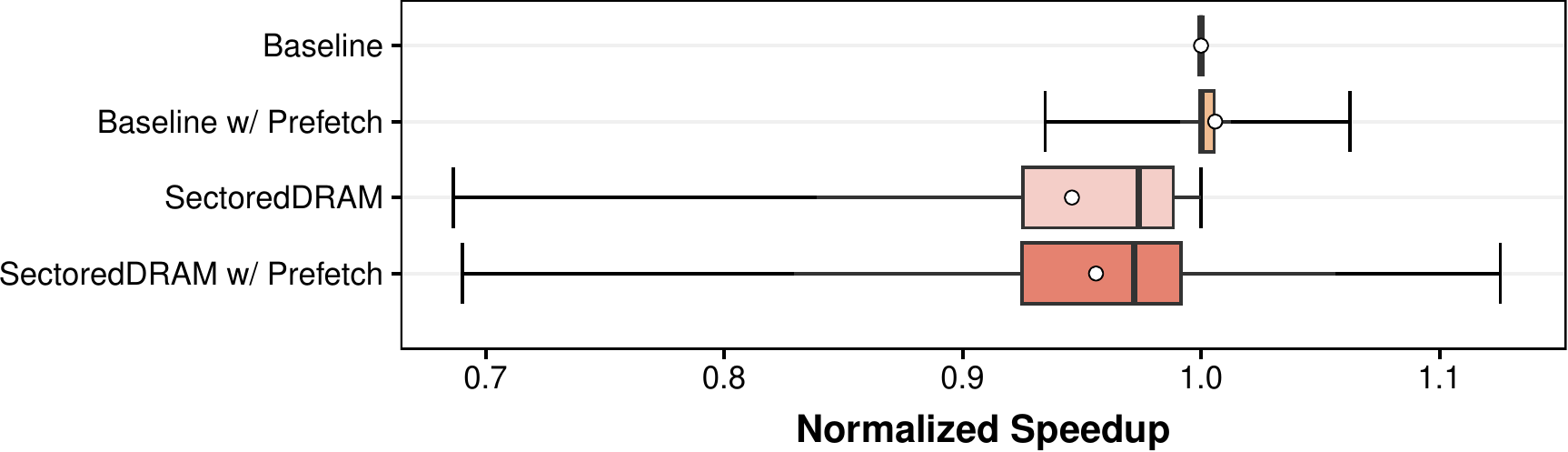}
    \caption{\tacorevcommon{Normalized speedup (x-axis) of all single-core workloads for Baseline and \X{}}\footref{footnote:box}}
    \label{fig:prefetching}
\end{figure*}

\tacorevcommon{We observe that \X{}-Prefetch improves system performance by 1.07\% (37.56\%) on average (at maximum) compared to \X{} without prefetching.\footnote{\tacorevcommon{Performance drop induced by the prefetcher can be curbed using prefetch throttling techniques~\cite{ebrahimi2009coordinated,ebrahimi2009techniques,ebrahimi2011prefetch}, e.g., Feedback Directed Prefetching~\cite{srinath2007feedback}.}} We conclude that the simple stride prefetcher design improves \X{}'s performance. We leave the design and analysis of a more sophisticated \X{} prefetcher for future work.}

\subsection{Finer-Granularity Sector Support}
\label{sec:discussion-finer-granularity-sectors}

{We design and evaluate \X{} with 8 sectors. Extending \X{} to support more sectors could enable higher energy and performance benefits but would require (i) transferring additional sector bits to DRAM \atbcr{1}{from the memory controller} and (ii) more DRAM circuit area to place additional sector latches.} 

{Our implementation allows us to transfer up to 14 sector bits with each $PRE$ command to DRAM (\cref{sec:sectored-activation-mechanism}). To transfer more than 14 sector bits, (i) DRAM command encoding could be extended with new signals to carry the sector bits (e.g., another signal/pin for \atbcr{1}{every} additional sector bit), (ii) a new DRAM command with enough space allocated in its encoding for sector bits could be implemented, or (iii) sector bits \atbcr{1}{for a single \XMO{} operation} could be \atbcr{1}{transferred to DRAM} over multiple $PRE$ commands.}

{\atbcr{1}{We evaluate t}he area required by additional sector latches \atbcr{1}{and find it to be} \omcr{2}{very small}. Implementing 8 more sector latches brings \X{}'s DRAM chip area overhead from 1.72\% to 1.78\%.}

\subsection{DRAM Error Correcting Codes (ECC)}

\revdel{\new{SECDED-ECC encodes a beat of the DRAM data transfer burst into a codeword by adding parity bits. The DRAM module is extended with additional chips to store the parity bits, and the memory channel is extended with additional DQ pins to transfer the parity bits. Since with \X{}, each DRAM access transfers at least one full codeword (64 bits data + 8 bits parity), ECC encoding or decoding operations can be correctly performed after each access.}}

\new{\revdel{Single-error-correcting double-error-detecting code (}SECDED-ECC\revdel{)}~\cite{mukherjee2008architecture} used in today's systems is naturally compatible with \X{}.} \new{Some systems} make use of more specialized, \microrev{Chipkill-like} ECC (e.g., single symbol error correction~\cite{amd2013sddc,yeleswarapu2020addressing, chen1996symbol}). \microrev{Sectored DRAM can easily support the single symbol error correction (SSC~\cite{amd2013sddc,yeleswarapu2020addressing, chen1996symbol}) scheme. In this scheme, the ECC codeword consists of 32 4-bit data symbols and four 4-bit ECC symbols with a total size of 144 bits. 128 data bits are encoded to form ECC symbols. The DRAM module consists of 16 x4 chips to store data symbols and 2 x4 chips to store ECC symbols. The module transmits 72 bits with each beat of the data\revdel{ transfer} burst and an ECC codeword is transmitted over two beats of a data\revdel{transfer} burst. To support the SSC scheme, \X{} can use burst lengths that are multiples of two\revdel{(i.e., 2, 4, 6, or 8)}, which allows the DRAM module to transmit whole ECC codewords \new{with every DRAM access}.} 

\Copy{R2/2}{\changev{\ref{q:r2q2}}\tacorevb{Recent DRAM chips implement on-die ECC which allows the DRAM chip to correct errors transparently from the memory controller~\cite{patelthesis,patel2019understanding,patel2020bit}. \X{} is compatible with on-die ECC schemes that operate at the granularity of a single sector (e.g., 8 bits). To develop an on-die ECC \atbcr{1}{scheme} for \X{}, existing on-die ECC schemes could be modified to operate at the granularity of a single sector or new ECC schemes could be developed. We leave such exploration for future work.}}

\subsection{Sector Cache Benefits}

\Copy{R4/10}{\changev{\ref{q:r4q10}}\tacorevd{We use sector caches to integrate \X{} into a full system. A comprehensive design space exploration for sector caches is out of the scope of this work (we refer the reader to~\cite{liptay1968structural, smith1987line, alpert1988performance, hill1984experimental, rothman2000sector,kadiyala1995dynamic,anderson1995two,rothman2002minerva,kuangchih1997on,kumar2012amoeba,pujara2006increasing}). Sector caches incur chip area costs but offer power and performance benefits beyond those that we demonstrate in our work (using \X{}). For example, powering off SRAM subarrays that contain invalid words in a cache block could save power \atbcr{1}{(e.g.,~\cite{pujara2006increasing})}, or filling up these invalid words with valid words from other cache blocks could increase effective cache capacity and improve system performance \atbcr{1}{(e.g.,~\cite{kumar2012amoeba})}}}.

\revdel{\subsection{\new{Burst Chop in DDRx}}

\new{Recent DDRx standards~\cite{jedec2007ddr3,jedec2017ddr4,jedec2020ddr5} support burst chop operation \gf{that} allows the DRAM chip to cut the data transfer burst's length in half. Burst chop can be used to enable Fine-DRAM-Access in a system \emph{without} any modifications to DRAM chips (although at an access granularity of only half of the cache block size).}

\new{We evaluate \X{} with only the burst chop operation (i.e., no \XMO{} and no \XMT{}) to see how system performance and DRAM energy are affected using high MPKI workload mixes. We observe that (i) \X{} reduces DRAM energy on average compared to the baseline system by 18\%, and (ii) \X{} reduces the average weighted speedup for the workload mixes by 5\%. Since \X{} with only the burst chop operation does not implement \XMO{}, these workloads cannot benefit from the reduction in $tFAW$. Moreover, sector misses result in extra memory accesses, increasing the average memory latency observed by these workloads. We conclude that \XMO{} and \XMT{} are critical for \X{} to enable high performance and low energy consumption.}}

\section{Related Work}
\X{} is the first low-cost and high-performance DRAM substrate that {alleviates the energy waste} \new{{on the DRAM bus,} by enabling (i) Fine-grained DRAM data {transfer} (Fine-DRAM-Transfer) and (ii) Fine-grained DRAM row activation (Fine-DRAM-Act)}. We extensively compare \X{} \omcr{2}{qualitatively and quantitatively} {with the most} relevant, low-cost, state-of-the-art fine-grain\gf{ed} DRAM architectures~\cite{cooper2010fine,zhang2014half,lee2017partial,ha2016improving,udipi2010rethinking} \gf{in \cref{sec:motivation:limitations} and \cref{sec:results-perf-energy}}. In this section, we discuss other related work\gf{s}.

\noindent
\textbf{Other Fine-Grain\gf{ed} activation mechanisms.} \new{Prior works~\cite{zhang2017enabling,alawneh2021dynamic,son2014microbank,oconnor2017fine,chatterjee2017architecting} propose other \gf{fine-grained DRAM} architectures. 
{These architectures require intrusive reorganization of} the DRAM array and/or {modifications to} the DRAM on-chip interconnect. \new{Some} works~\cite{chatterjee2017architecting, oconnor2017fine, son2014microbank} target higher-bandwidth DRAM standards and offer significant performance and activation energy improvements. \cite{zhang2017enabling}~develops a new interconnect that serializes \new{data} from multiple partially activated banks. \cite{alawneh2021dynamic}~divides DRAM mats into submats by adding more helper flip-flops to mitigate the throughput loss of fine-grain\gf{ed} activation. These works do \emph{not} reduce the energy wasted on the memory channel by \atbcr{1}{avoiding the transfer of unused words}, which \X{} does at low chip area cost.}

\noindent
\textbf{DRAM-module-level Fine-DRAM-{Transfer}.} A class of prior work~\cite{yoon2011adaptive,yoon2012dynamic,ahn2009multicore,zheng2008minirank,brewer2010instruction,ware2006improving} propose\gf{s} new DRAM module designs (e.g., subranked DIMMs~\cite{ahn2009multicore,yoon2011adaptive,yoon2012dynamic,ware2006improving}) that allow independent operation of each DRAM chip \gf{in a DRAM module} \gf{(\cref{sec:background:dramorg})} \gf{in order to} \new{implement Fine-DRAM-{Transfer}}.
\microrev{From a system standpoint, \X{} is a more practical mechanism to implement compared to module-level mechanisms \new{because \X{} requires no modifications to the physical DRAM interface}. Similar to \X{}, \new{these} mechanisms (e.g., DGMS~\cite{yoon2012dynamic}) require modifications to DRAM (i.e., modifications to the DRAM module in DGMS and to the DRAM chip in \X{}) and the processor. However, on top of these modifications, module-level mechanisms also require modifications to the physical DRAM interface \new{(e.g., three additional pins to select one of the eight chips in a rank~\cite{ahn2009multicore})}, thus making module-level mechanisms incompatible with \atbcr{1}{current} industry standards (i.e., JEDEC specifications~\cite{jedec2017ddr4}). In contrast, \X{} does \emph{not} require modifications to the physical DRAM interface, and thus, \X{} chips comply with \atbcr{1}{existing DRAM} industry standards and specifications.} 

\new{We quantitatively evaluate a subranked DIMM design (DGMS~\cite{yoon2012dynamic}) that can be implemented with minimal modifications to the physical DRAM interface. This design can operate subranks independently and each subrank can receive one DRAM command per DRAM command bus cycle (i.e., 1x ABUS scheme~\cite{yoon2011adaptive}). We find that this design \emph{reduces} system performance for the high MPKI workload mixes, causing a 23\% reduction in weighted speedup on average. Even though the subranked DIMM allows requests to be served from different subranks in parallel, the DRAM command bus bandwidth is insufficient to allow timely scheduling of these requests to different subranks~\cite{yoon2012dynamic} (i.e., the command bus becomes the bottleneck). The DRAM command bus bandwidth can be increased to enable higher-performance subranked DIMM designs. However, this comes at additional hardware cost and modifications to the physical DRAM interface~\cite{yoon2012dynamic}. In contrast, \X{} improves the weighted speedup for the same set of workloads by 17\% on average and requires no modifications to the physical DRAM interface.}

\section{Conclusion}
\new{We design\atbcr{1}{ed} a new, high-throughput, energy-efficient, and practical fine-grained DRAM architecture, \X{}. Compared to prior fine-grained DRAM architectures, our design significantly improves both system energy and performance. It does so by eliminating 
(i) the energy waste \gf{caused by} transferring unused words between the processor and DRAM, and
(ii) the energy spent on activating DRAM cells that are not accessed by memory requests. 
Activating a smaller number of cells allows the memory controller to serve memory requests with lower memory \gf{access} latency.
While effective at improving both system energy and performance, \X{} is also practical and can be implemented at low hardware cost.}

\section*{Acknowledgements}
\atbcr{1}{We thank the anonymous reviewers of MICRO 2022, HPCA 2023, and TACO for their feedback. 
We thank the SAFARI Research Group members for providing a stimulating intellectual environment. We acknowledge
the generous gifts from our industrial partners{; including} Google, Huawei, Intel, \atbcr{2}{and} Microsoft{.} This work is supported in part by the Semiconductor Research Corporation{,} the ETH Future Computing Laboratory{, and the {AI Chip Center for Emerging Smart Systems (ACCESS)}}.}

\balance
\begin{spacing}{0.75}
\begin{footnotesize}
\bibliographystyle{IEEEtranS}
\bibliography{refs}
\end{footnotesize}
\end{spacing}

\end{document}